\documentclass[aps,pra,superscriptaddress,showpacs,twocolumn]{revtex4-1}
\usepackage{bm}
\usepackage{amsmath}
\usepackage{amssymb}
\usepackage{slashed}
\usepackage{float}
\allowdisplaybreaks
\usepackage{subcaption}
\usepackage{dcolumn}
\usepackage{graphicx}
\newcolumntype{.}{D{x}{}{-1}}

\newcommand*{\centt}[1]{\multicolumn{1}{c}{#1}}
\newcolumntype{w}[1]{D{.}{.}{#1}}
\newcolumntype{L}{>{$}l<{$}}

\usepackage{graphicx}
\usepackage{xcolor}
\usepackage{nicefrac}

\newcommand{\lbr}{\langle}
\newcommand{\rbr}{\rangle}

\newcommand{\qsla}{q\hspace{-0.45em}/}

\begin{document}

\title{Radiative $\bm{\alpha^7m}$ QED contribution to the helium Lamb shift}

\author{Vojt\v{e}ch Patk\'o\v{s}}
\affiliation{Faculty of Mathematics and Physics, Charles University,  Ke Karlovu 3, 121 16 Prague
2, Czech Republic}

\author{Vladimir A. Yerokhin}
\affiliation{Center for Advanced Studies, Peter the Great St.~Petersburg Polytechnic University,
Polytekhnicheskaya 29, 195251 St.~Petersburg, Russia}

\author{Krzysztof Pachucki}
\affiliation{Faculty of Physics, University of Warsaw,
             Pasteura 5, 02-093 Warsaw, Poland}

\begin{abstract}

We present a derivation of the last unknown part of the $\alpha^7m$ contribution to the Lamb
shift of a two-electron atom, induced by the radiative QED effects beyond the Bethe logarithm.
This derivation is performed in the framework of nonrelativistic quantum electrodynamics and is
valid for the triplet (spin $S = 1$) atomic states. The obtained formulas are free from any
divergences and are suitable for a numerical evaluation. This opens a way for a complete numerical 
calculation of the $\alpha^7m$ QED effects in helium, which will allow an accurate determination 
of the nuclear charge radius from measurements of helium transition frequencies.

\end{abstract}

\maketitle

\section{Introduction}

The helium atom has been extensively studied since the very advent of quantum mechanics as the
prototypical three-body problem and an ideal testing ground for various theoretical approaches
describing many-electron atoms. Modern theoretical calculations of helium spectra
\cite{pachucki:17:heSummary} approach the level of accuracy previously achievable only for the
simplest case of the hydrogen atom, with a potential for discovery of new effects beyond the Standard
Model. Among the major topics of current interest in the helium atom are possibilities to obtain an
'atomic physics' value for the fine structure constant $\alpha =
\nicefrac{e^2}{(4\pi\epsilon_0\hbar c)}$ \cite{pachucki:10:hefs,kato:18} and to provide a novel
method for the determination of the nuclear charge radius \cite{pachucki:15:jpcrd,zheng:17}.

Up to now,  theoretical calculations of the helium Lamb shift allowed accurate determinations
only for the {\em differences} of the nuclear charge radii of two isotopes
\cite{patkos:17:singlet}. In order to determine the {\em absolute} value of the nuclear radius
with a precision comparable to what is expected from muonic helium \cite{pohl:16:leap}, one
needs to perform a complete calculation of the $\alpha^7m$ QED effects, which is a very
challenging task.

The $\alpha^7m$ QED effects were first calculated for hydrogen in
Refs.~\cite{pachucki:93,pachucki:94,eides:95:pra} and recently generalized for the two-center
hydrogenic problem \cite{korobov:14}. For helium, the $\alpha^7m$ QED effects were investigated
in the context of the fine-structure splitting
\cite{pachucki:06:prl:he,pachucki:09:hefs,pachucki:10:hefs}, which offered extensive
simplifications to the underlying theory.

Several years ago we started a project of calculating the $\alpha^7m$ QED effects in the Lamb
shift of helium. The first aim of the project is the triplet (spin $S = 1$) states, whose theory
is simplified by the absence of the so-called contact operators $\propto
\delta(\vec{r}_1-\vec{r}_2)$. Intermediate results were published in
Refs.~\cite{yerokhin:18:betherel,patkos:20}. In the present work, we complete the derivation of
the $\alpha^7m$ contribution, presenting formulas for the last missing part. More specifically, the
$\alpha^7m$ contribution to the Lamb shift are represented as a sum of three parts,
\begin{align}\label{eq:0}
  E^{(7)} =   E_L^{(7)} + E^{(7)}_{\rm exch} + E^{(7)}_{\rm rad}\,,
\end{align}
where $E_L^{(7)}$ is the relativistic correction to the so-called Bethe logarithm, $E^{(7)}_{\rm
exch}$ is the part induced by the electron-electron and electron-nucleus photon exchange, and
$E^{(7)}_{\rm rad}$ is induced by the radiative QED effects beyond the Bethe logarithm. The
relativistic correction to the Bethe logarithm was calculated in
Ref.~\cite{yerokhin:18:betherel}, whereas the photon-exchange contribution was derived by us in
Ref.~\cite{patkos:20}. The goal of the present investigation is to perform a derivation of the
last missing part in Eq.~(\ref{eq:0}), thus obtaining the total set of formulas for the
$\alpha^7m$ contribution for the triplet states of a two-electron atom. The numerical evaluation
of the obtained formulas is left for the forthcoming investigation.

\section{Basic approach}

The radiative ${\alpha^7\,m}$ contribution to the Lamb shift is represented by a sum of the
first-order and the second-order perturbation corrections,
\begin{align}\label{eq:1}
  E^{(7)}_{\rm rad} = &\
  \lbr H^{(7)}_{\rm rad} \rbr
  + 2\,\big<  H^{(4)}  \, \frac{1}{(E_0-H_0)'} \, H^{(5)}_{\rm rad} \big>
     \,.
\end{align}
Here, $H^{(7)}_{\rm rad}$ and $H^{(5)}_{\rm rad}$ are the effective Hamiltonians induced by the
radiative effects of order $\alpha^7m$ and $\alpha^5m$, respectively, $H^{(4)}$ is the Breit
Hamiltonian (of order $\alpha^4m$), and $H_0$ and $E_0$ are the nonrelativistic Hamiltonian and the
corresponding eigenvalue, respectively.

$E^{(7)}_{\rm rad}$ consists of contributions coming from the one-loop self-energy (SE), the
one-loop vacuum polarization (VP), and the two-loop (rad2) and three-loop QED effects (rad3),
\begin{align}
E^{(7)}_{\rm rad} = &\  E^{(7)}_{\rm SE} + E^{(7)}_{\rm VP}+ E^{(7)}_{\rm rad2} + E^{(7)}_{\rm rad3}.
\end{align}
The main part of the present investigation will be devoted to the evaluation of the one-loop
self-energy correction, which is by far the most difficult part. The corresponding derivation is
presented in Secs.~\ref{sec:hamrel}-\ref{sec:SE}. The one-loop vacuum polarization is calculated
in Sec.~\ref{sec:VP}, whereas the two- and three-loop radiative effects are obtained in
Sec.~\ref{sec:multiloop}.

One of the major problems encountered in deriving formulas for higher-order QED effects is
connected with numerous divergences appearing on intermediate stages of the calculation. In order
to systematically handle these divergences, we use dimensionally regularized nonrelativistic
quantum electrodynamics (NRQED), with the dimension of the space-time $D= 4-2\,\epsilon$ and the
dimension of space $d=3-2\,\epsilon$. The parameter $\epsilon$ is considered as small, but only
on the level of matrix elements, where the analytic continuation to a noninteger spatial
dimension is allowed. The final results will be expanded in small $\epsilon$, and singular
contributions $\propto 1/\epsilon$ will be canceled algebraically in momentum space. Subsequently,
the results will be transformed into the coordinate representation, where they can be
calculated numerically. The foundation of the dimensionally regularized NRQED in the context of
the hydrogen Lamb shift was laid in Ref.~\cite{pineda:98}, but our approach differs in many
details.

In our derivation of the radiative ${\alpha^7\,m}$ correction for helium we will rely on the fact
that the corresponding correction is known for hydrogen-like atoms (see, e.g., a review
\cite{yerokhin:18:hydr}). With this in mind, in our derivation we will repeatedly drop the
first-order terms containing purely the electron-nucleus Dirac $\delta$-function. (We still have
to keep terms in which the electron-nucleus $\delta$-function is combined with the potential,
energy or momenta, however.) This procedure will simplify the derivation enormously and the
omitted $\delta$-like terms will be later restored by matching the known hydrogenic results. Our
calculation will be performed in the Coulomb gauge, unless explicitly specified otherwise.

When working in the dimensionally regularized NRQED, we need generalizations of basic operators
into the extended number of dimensions, shortly summarized below. The momentum-space
representation of the photon propagator preserves its form in $d$ dimensions, namely,
$g_{\mu\nu}/k^2$. The surface area of a $d$-dimensional unit sphere is
\begin{equation}
\Omega_d = \frac{2\,\pi^{d/2}}{\Gamma(d/2)}\,.
\end{equation}
The electron-nucleus Coulomb interaction becomes
\begin{align}
V_C(r) &= -Z\,e^2\,\int\frac{d^d k}{(2\,\pi)^d}\,
\frac{e^{{\rm i}\,\vec k\cdot \vec r}}{k^2}
\nonumber\\
&= -\frac{Z\,e^2}{4\,\pi\,r^{1-2\,\epsilon}}\,
\left[(4\,\pi)^\epsilon\,
\frac{\Gamma(1-2\,\epsilon)}{\Gamma(1-\epsilon)}\right]
\equiv -  \left[ \frac{Z\alpha}{r}\right]_{\epsilon}\,. \label{04}
\end{align}
Here and in what follows, $[ X ]_{\epsilon}$ will denote the $d$-dimensional generalization of
the $d=3$ operator $X$.
The $d$-dimensional nonrelativistic Hamiltonian of a two-electron atom is given by
\begin{equation}
H_0 = \sum_{a = 1,2}\frac{\vec{p}_a^{\;2}}{2\,m}  + V\,,
\label{06}
\end{equation}
where
\begin{equation}\label{eq:V}
  V = -\bigg[\frac{Z\,\alpha}{r_1}\bigg]_\epsilon -\bigg[\frac{Z\,\alpha}{r_2}\bigg]_\epsilon
  + \bigg[\frac{\alpha}{r}\bigg]_\epsilon\,,
\end{equation}
and $\vec{r} = \vec{r}_{1}-\vec{r}_{2}$. The $d$-dimensional generalization of the Breit
Hamiltonian for a two-electron atom is
\begin{align}\label{eq:hbreit}
 H^{(4)}&\  = H'^{(4)} + H''^{(4)}\,, \\
\label{eq:hbreit1}
 H'^{(4)}&\ = - \frac{\pi\,\alpha}{m^2}\,\delta^d(r)
+ \sum_{a=1,2}\biggl\{ -\frac{p_a^4}{8\,m^3} +\frac{\pi\,Z\alpha}{2\,m^2}\,\delta^d(r_a)\biggr\}
\nonumber \\ &\
-\frac{\alpha}{2\,m^2}\,p_1^i\,
\biggl[\frac{\delta^{ij}}{r}+\frac{r^i\,r^j}{r^3}
\biggr]_\epsilon\,p_2^j
- \frac{\pi\,\alpha}{d\,m^2}\,\sigma_1^{ij}\,\sigma_2^{ij}\,\delta^d(r)\,,
  \\
 H''^{(4)} &\ =
 \sum_{a=1,2} \frac{1}{4\,m^2}\,\sigma_a^{ij}\,\bigl(\nabla^i_a V\bigr) p_a^j
 \nonumber \\ &\
+\frac{1}{4\,m^2}\,\sigma_1^{ik}\,\sigma_2^{jk}\,
\left(\nabla^i\,\nabla^j-\frac{\delta^{ij}}{d}\,\nabla^2\right)\,
\left[\frac{\alpha}{r}\right]_\epsilon \nonumber \\ &\
-\frac{1}{2\,m^2}\,\biggl(\sigma_1^{ij}\,\nabla^i\left[\frac{\alpha}{r}\right]_\epsilon\,p^j_2
-\sigma_2^{ij}\,\nabla^i\left[\frac{\alpha}{r}\right]_\epsilon\,p^j_1 \biggr)\,,
\label{eq:hbreit2}
\end{align}
where  $\delta^{d}(r)$ is the Dirac $\delta$-function in $d$ dimensions, $\sigma^{ij} = 1/(2{\rm
i})[\sigma^i,\sigma^j]$, and $\sigma^i$ are the Pauli matrices. In $d=3$ spatial dimensions, the
matrices $\sigma^{ij}$ reduce to $\sigma^{ij}=\epsilon^{ijk} \, \sigma^k$.  The $d$-dimensional
generalization of the algebra of the Pauli matrices is summarized in Appendix B of
Ref.~\cite{patkos:20}.

\section{NRQED Hamiltonian}
\label{sec:hamrel}

We now turn to the derivation of the NRQED Hamiltonian incorporating relativistic corrections and a
class of radiative corrections that comes from the high photon momentum, in $d$ dimensions. This Hamiltonian can be obtained from
the Dirac Hamiltonian by means of the Foldy-Wouthuysen transformation. In order to incorporate a high-energy
part of the radiative corrections, we start with the Dirac Hamiltonian modified by the
electromagnetic form factors $F_1$ and $F_2$,
 \begin{align}
\label{HDirac}
H_D =&\ \vec{\alpha} \cdot
\left[\vec{p} - e \, F_1(\vec \nabla^2) \, \vec{A}\right] + \beta\,m
+ e\,F_1(\vec \nabla^2) \, A_0
\nonumber\\
& + F_2(\vec \nabla^2) \, \frac{e}{2\,m} \, \left({\rm i}\,\vec{\gamma} \cdot
\vec{E} - \frac{\beta}{2} \, \Sigma^{ij}\,B^{ij} \right)\,,
\end{align}
where $ B^{ij} = \nabla^i\,A^j-\nabla^j\,A^i\,$, $\nabla^i \equiv \nabla_i =
\partial/\partial x^i$, and $\Sigma^{ij} = \frac{\rm i}{2}\,[\gamma^i,\gamma^j]$.
Formulas for the electromagnetic form factors $F_{1}$ and $F_2$ are summarized in
Appendix~\ref{app:ffactors}. Introducing the Foldy-Wouthuysen transformation defined by the
operator $S$,
\begin{align}
S =& -\frac{\rm i}{2\,m}\,\left\{
\beta\,\vec\alpha\cdot\vec\pi
-\frac{1}{3\,m^2}\,\beta\,(\vec\alpha\cdot\vec\pi)^3\right.
\nonumber\\
& \left. +\frac{e(1+\kappa)}{2\,m}\,{\rm i}\,\vec\alpha\cdot\vec E
-\frac{e\,\kappa}{8\,m^2}\,
[\vec\alpha\cdot\vec\pi, \beta\,\Sigma^{ij}\,B^{ij}]\right\}\,,
\label{13}
\end{align}
where  $\kappa \equiv F_2(0)\approx \alpha/(2\,\pi)$ is the electron anomalous magnetic moment,
the transformed Hamiltonian is obtained as
\begin{equation}
\label{trafo1}
H_{FW} = e^{{\rm i}\,S}\,(H_D-{\rm i}\,\partial_t)\,e^{-{\rm i}\,S}\,.
\end{equation}

We will split the transformed Hamiltonian into two parts, $\delta H_1$ and $\delta H_2$, $H_{FW}
= \delta H_1 + \delta H_2$, where $\delta H_1$ contains the form factors only in the form of the
anomalous magnetic moment $\kappa$, and $\delta H_2$ is the remainder.
\begin{widetext}
The result is
\begin{align}
\label{trafo3}
 \delta H_1 = &\ \frac{\vec \pi^{2}}{2\,m}  +
e A^0 -
\frac{e}{4\,m}\,(1+\kappa)\,\sigma^{ij}\,B^{ij}
-\frac{\vec \pi^{\;4}}{8\,m^3}
-\frac{e}{8\,m^2}\,(1+2\,\kappa)\,
\left[\vec\nabla\cdot\vec E+\sigma^{ij}\,\{E^i,\pi^j\}\right]
\nonumber \\
& +\frac{e}{16\,m^3}\left[\left(1+\frac\kappa2\right)\,\{p^2\,,\,\sigma^{ij}\,B^{ij}\}+
4\,\kappa\,p^k\,\sigma^{ki}\,B^{ij}\,p^j + \kappa\,\vec p\,\,\sigma\cdot B\,\vec p\right]
+\frac{\vec{p}^{\,6}}{16\,m^5}
+\frac{(\kappa^2+\kappa+1)}{8\,m^3}\,e^2\,\vec E^2
\nonumber \\ &
+\frac{3+4\,\kappa}{64\,m^4}e\,
\bigl\{\vec p^{\;2},\,\vec\nabla\cdot\vec E+\sigma^{ij}\,\{E^i,p^j\}\bigr\}
+ \frac{5\,{\rm i}\,e}{128 m^4}\,[p^2, \{\vec{p}\,,\,\vec{E}\}  ]
- \frac{e\,\kappa}{16 m^3} \,
         \{\pi^k, \partial_{t}E^k + \nabla^i\,B^{ik}\} + \ldots,\\
\delta H_2 = &\
e\Big[F'_1(0)\,\Delta + \frac12F''_1(0)\,\Delta^2\Big] A^0
-\frac{e}{m} \Big(F'_1(0)\vec p\cdot\Delta\vec A
+\frac{1}{4}\, \big[F'_1(0)+F'_2(0)\big]\,\sigma^{ij}\,\Delta B^{ij}\Big)
\nonumber\\&
 -\frac{e}{8\,m^4}\,\big[F'_1(0)+2\,F'_2(0)\big]\,\Delta\,\vec\nabla\cdot\vec E + \ldots,
\label{trafo3b}
\end{align}
\end{widetext}
where $\Delta = \vec\nabla^2$, $\{X,Y\} \equiv X\,Y+Y\,X$, and $\vec \pi = \vec p-e\,\vec A$. The
ellipsis denotes higher-order terms and spin-dependent terms that do not contribute to the
centroid energies.

It should be noted that the neglect of $\partial/\partial x^0$ within the $F_1$ and $F_2$ form
factors is not valid for the electron-electron interactions. In this case we will calculate
effective operators from the one-photon exchange scattering amplitude in the Feynman gauge.
Moreover, there is  an additional radiative correction to the NRQED Hamiltonian that is
not accounted for by the $F_1$ and $F_2$ form factors. It is represented by an effective local
operator quadratic in the field strengths. This operator is derived separately by evaluating a
low-energy limit of the electron scattering amplitude off the Coulomb field. It was calculated in
Ref. \cite{jentschura:05:sese} to be
\begin{equation}
\label{deltaH}
\delta H_3 =
\frac{e^2}{m^3}\,\vec E^{\;2}\,\chi\,,
\end{equation}
where $E$ is an electric field, and the function $\chi$ is given
by
\begin{equation}
\chi = \frac{\alpha}{\pi}\bigg(\frac16-\frac{1}{3\epsilon}\bigg)\,,
\end{equation}
where we included only the one-loop part. We follow throughout this work the convention of Ref. \cite{jentschura:05:sese}, that a common factor of \mbox{$\big[(4\pi)^\epsilon\,
\Gamma(1+\epsilon)\big]$} for each loop integration is pulled out from all matrix elements.

We thus obtain the effective NRQED Hamiltonian as
\begin{equation} \label{eq:hrel}
H_{\rm nrqed} = \delta H_1 + \delta H_2 + \delta H_3\,.
\end{equation}
Individual operators from $H_{\rm nrqed}$ induce numerous contributions to the one-loop electron
self-energy, evaluated term by term in Sec.~\ref{sec:EH}.

\section{Effective $\bm{\alpha^5m}$ Hamiltonian}

We now derive the self-energy part of the effective $\alpha^5\,m$ operator $H^{(5)}_{\rm rad}$,
which is required for evaluation of the second-order contribution in Eq.~(\ref{eq:1}). Despite
the fact that the $\alpha^5\,m$ correction is well-known  in principle \cite{araki:57,sucher:58},
we here need a $d$-dimensional generalization of the corresponding effective operator. From now
on we will make the simplification of setting the electron mass equal to one, $m=1$.

The leading-order self-energy correction $E^{(5)}_{\rm SE}$ can be written as a sum of low-energy
and high-energy contributions,
\begin{equation}\label{eq:19}
E^{(5)}_{\rm SE} = E_{L0} + E_{H0}.
\end{equation}
The low-energy part comes from the radiative photon momenta of the order $k\propto \alpha^2\,m$.
It is represented by the one-loop self-energy contribution in the leading nonrelativistic-dipole
approximation,
\begin{eqnarray}\label{eq:L:0}
E_{L0} &=& e^2\int \frac{d^dk}{(2\pi)^d2k}\,
 \delta_\perp^{ij}(k)
 \nonumber \\ && \times
\,\Big\langle\phi\Big|p_1^i\, \frac{1}{E_0-H_0-k}\,p_1^j\Big|\phi\Big\rangle
+(1\leftrightarrow2)\,,
\end{eqnarray}
where $\delta_\perp^{ij}(k) = \delta^{ij} - k^ik^j/k^2$. The integration over $|k|$ is
split into two parts, $\int_0^\infty = \int_0^\Lambda + \int_\Lambda^\infty$, leading to the
separation
\begin{equation}\label{22}
E_{L0} = E_L^{(5)} + E^\Lambda_{L0}\,,
\end{equation}
where the first term $E_L^{(5)}$ is the electron self-energy part of the Bethe logarithm in three dimensions,
\begin{eqnarray}\label{eq:22}
E_L^{(5)} &=& \frac{2e^2}{3}\int_0^\Lambda \frac{d^3k}{(2\pi)^32k}\,
\,\Big\langle\phi\Big|p_1^i\, \frac{1}{E_0-H_0-k}\,p_1^i\Big|\phi\Big\rangle
 \nonumber \\ &&
+(1\leftrightarrow2)\,,
\end{eqnarray}
while the second terms is the remainder to be calculated in $d$-dimensions.
It is assumed that
\begin{equation}\label{eq:23}
\Lambda = \alpha^2\lambda\,,
\end{equation}
with arbitrary large $\lambda$. Namely, it is assumed that the limit $\epsilon\rightarrow 0$
is performed as the  first and $\lambda\rightarrow\infty$ as the second one.
So, we obtain
\begin{eqnarray}\label{21}
E_L^{(5)} &=&-\frac{2\alpha}{3\pi}\Big\langle\phi\Big|p_1^i\, (H_0-E_0)
\ln\bigg[\frac{(H_0-E_0)}{\alpha^2}\bigg]\,p_1^i\Big|\phi\Big\rangle
 \nonumber\\
&&
+\frac{\alpha\ln\lambda}{3\pi}\nabla_1^2 V+(1\leftrightarrow2)\,.
\end{eqnarray}
Here, the second term depends on $\lambda$, but it will cancel out in the sum in Eq.~(\ref{22}). The remainder $E^\Lambda_{L0}$ is evaluated as
\begin{eqnarray}\label{eq:L:0:lam}
E^\Lambda_{L0} &=& e^2\int_\Lambda^\infty \frac{d^dk}{(2\pi)^d2k}\,
 \delta_\perp^{ij}(k)
 \nonumber \\ && \times
\,\bigg\langle\phi\bigg|p_1^i\, \frac{1}{E_0-H_0-k}\,p_1^j\bigg|\phi\bigg\rangle
+(1\leftrightarrow2)\nonumber\\
 &=& \frac{e^2}{4}\frac{d-1}{d}\int_\Lambda^\infty \frac{d^dk}{(2\pi)^dk^3}\,
 \nonumber \\ && \times
\,\bigg\langle\phi\bigg|\big[p_1^i, \big[H_0-E_0,p_1^i\big]\big]\bigg|\phi\bigg\rangle
+(1\leftrightarrow2)\,.
\end{eqnarray}
Performing the integration with respect to $k$ we obtain
\begin{align}\label{eq:26}
E_{L0}^\Lambda &\ =\frac{\alpha}{\pi}\mathcal{I}_\epsilon\,\big(\nabla_1^2 V + \nabla_2^2 V\big)
\,,
\end{align}
where
\begin{eqnarray}
\mathcal{I}_\epsilon &=& \pi^2\frac{d-1}{d}\int_\Lambda^\infty \frac{d^dk}{(2\pi)^dk^3}\nonumber\\
&=&
-\frac{2^{-d} (d-1) \pi ^{2-\frac{d}{2}} \left[\alpha^2\lambda\right]^{d-3}}{(d-3) \Gamma
   \left(\frac{d}{2}+1\right)}
	= \frac{5}{18}+\frac16\ln\Lambda_\epsilon + O(\epsilon)\nonumber\\
\end{eqnarray}
and
\begin{equation}
\ln{\Lambda_{\epsilon}} = \frac{1}{\epsilon}+2\ln(\alpha^{-2})-2\ln(2\lambda)\,.
\end{equation}

The high-energy contribution, $E_{H0}$ in Eq.~(\ref{eq:19}), is induced by the region where the
momentum of the radiative photon is of the order of the electron mass $k\propto m$. $E_{H0}$ is
conveniently split into two parts,
\begin{equation}
E_{H0}= E_{H0}^A+E_{H0}^B.
\end{equation}
The first part comes from the exchange of a transverse photon with one vertex
$-\frac{e}{4}\kappa\,\sigma\cdot B$. It is calculated in the nonretardation approximation, with
the result
\begin{widetext}
\begin{align}\label{eq:29}
E_{H0}^A =&\  -e^2\kappa \int \frac{d^dk}{(2\pi)^d\,2k^2}\delta_\perp^{ij}(k)
\langle\phi|\biggl(p_1^i+\frac{1}{2}\sigma_1^{ki}\nabla_1^k\biggr)e^{{\rm i}\vec{k}\cdot\vec{r}_1}
\biggl(\frac{1}{2}\sigma_2^{lj}\nabla_2^l\biggr)e^{-{\rm i}\vec{k}\cdot\vec{r}_2}|\phi\rangle
+\textrm{h.c.}+ (1\leftrightarrow2)\nonumber\\
=&\ \alpha\kappa\,
\bigg<
 \frac14\sigma_1^{ik}\sigma_2^{jk}\bigg(\nabla^i\nabla^j-\frac{\delta^{ij}}{d}\nabla^2\bigg)
\bigg[\frac{1}{r}\bigg]_\epsilon-\frac{\pi}{d}\sigma_1\cdot\sigma_2\,\delta^d(r)
+\frac12\sigma_2^{ij}\nabla_1^i\bigg[\frac{1}{r}\bigg]_\epsilon\,p_1^j
 \bigg>
+(1\leftrightarrow2).
\end{align}
The remaining high-energy contribution comes from the exchange of a Coulomb photon with the
vertices $eA^0$ and $-\frac{e}{4}\kappa \vec\nabla\cdot\vec E
-\frac{e}{4}\kappa\,\sigma^{ij}\{E^i,p^j\}
 +F'_1(0)\,\vec\nabla^2 eA^0$.
The result  is
\begin{equation} \label{eq:30}
E_{H0}^B = \Big< \Big[F'_1(0)+\frac{1}{4}\kappa\Big] \nabla^2_1 V
+\frac{1}{4}\kappa\,\sigma_1^{ij}\{\nabla_1^i V,p_1^j\}\Big> + (1\leftrightarrow2)\,.
\end{equation}
Here $F'_1(0)$ contains the singular part $\propto1/\epsilon$, see Appendix \ref{app:ffactors}. This
singularity will cancel exactly with that in $\ln\Lambda_\epsilon$, so that the final result for
$E^{(5)}_{\rm SE}$ is finite. We thus obtain
\begin{equation}\label{eq:31}
E^{(5)}_{\rm SE} = E_{L}^{(5)} + \lbr H^{(5)}_{\rm SE} \rbr\,,
\end{equation}
where the self-energy Hamiltonian $H^{(5)}_{\rm SE}= H'^{(5)}_{\rm SE}  + H''^{(5)}_{\rm SE}$ is
given by
\begin{align}\label{eq:hma5a}
H'^{(5)}_{\rm SE} = &\
\frac{\alpha}{\pi }\bigg(\mathcal{I}_\epsilon-\frac{1}{6\epsilon}\bigg)
\big(\nabla_1^2 V + \nabla_2^2 V\big)
-\alpha\frac{2\pi\kappa}{d}\sigma_1\cdot\sigma_2\,\delta^d(r)\nonumber\\
\approx&\
\frac{\alpha}{\pi }\bigg[\frac{5}{18}+\frac13\ln(\alpha^{-2})-\frac13\ln(2\lambda)\bigg]
\big(\nabla_1^2 V + \nabla_2^2 V\big)
-\alpha\frac{2\pi\kappa}{3}\sigma_1\cdot\sigma_2\,\delta^d(r)\,,\\
H''^{(5)}_{\rm SE} =&\  \alpha\frac{\kappa}{2}\bigg\{\bigg[\vec\sigma_2\cdot\bigg(\vec\nabla_1\frac{1}{r}\bigg)\times\,\vec p_1
+\vec\sigma_1\cdot\vec\nabla_1 V\times \vec p_1 + (1\leftrightarrow2)\bigg]
-\sigma_1^{i}\sigma_2^{j}\bigg(\nabla^i\nabla^j-\frac{\delta^{ij}}{3}\nabla^2\bigg)
\frac{1}{r}\bigg\}\,.
\label{eq:hma5b}
\end{align}
\end{widetext}
Here we used $d = 3$ for $H''^{(5)}_{\rm SE}$ because the corresponding $\alpha^7m$ second-order
contribution will not contain any singularities. Note that the second line in
Eq.~(\ref{eq:hma5a}) holds only for the $\alpha^5m$ correction. For the $\alpha^7m$ contribution
coming from the second-order perturbation correction with $H^{(5)}_{\rm SE}$, we will have to
keep $\mathcal{I}_\epsilon$ in the closed form because of a contribution from the linear in
$\epsilon$ terms.
We also note that both $E_{L}^{(5)}$ and $H^{(5)}_{\rm SE}$ are free of any divergencies but they
depend on the cutoff parameter $\lambda$, which cancels out in their sum.

\section{Second-order $\bm{\alpha^7m}$ self-energy correction}
We now turn to the derivation of the second-order $\alpha^7m$ self-energy contribution
represented by the second term in Eq.~(\ref{eq:1}),

\begin{align}
E^{(7)}_{\rm sec, SE} = 2\,\langle H^{(4)}\,\frac{1}{(E_0-H_0)'}\,H^{(5)}_{\rm SE}\rangle\,,
\end{align}
where $H^{(4)}$ is the Breit Hamiltonian  and $H^{(5)}_{\rm SE}$ was derived in the previous
section. It is convenient to split $E^{(7)}_{\rm sec, SE}$ into the spin-dependent and
spin-independent parts,
\begin{align}
E^{(7)}_{\rm sec, SE} = &\
2\,\langle H'^{(4)}\,\frac{1}{(E_0-H_0)'}\,H'^{(5)}_{\rm SE}\rangle
 \nonumber \\ &
+2\,\langle H''^{(4)}\,\frac{1}{(E_0-H_0)'}\,H''^{(5)}_{\rm SE}\rangle\,,
\end{align}
where $H'^{(4)}$ and $H''^{(4)}$ are given by Eqs.~(\ref{eq:hbreit1}) and (\ref{eq:hbreit2}),
respectively, and $H'^{(5)}_{\rm SE}$ and $H''^{(5)}_{\rm SE}$ by Eqs.~(\ref{eq:hma5a}) and
(\ref{eq:hma5b}), respectively. The spin-dependent second-order contribution with $H''^{(5)}_{\rm
SE}$ is finite and can be calculated numerically as it stands. The contribution induced by
$H'^{(5)}_{\rm SE}$, however, is divergent and needs to be regularized and transformed in order
to move all $1/\epsilon$ singularities into the first-order terms where they can be canceled.

Omitting terms $\propto \delta^d(r)$ vanishing for triplet states, we rewrite the first part of
the $\alpha^5$ Hamiltonian as
\begin{align}
H'^{(5)}_{\rm SE} =&\  \frac{\alpha}{\pi }\bigg(\mathcal{I}_\epsilon-\frac{1}{6\epsilon}\bigg)\big[\vec P,\big[V,\vec
P\big]\big]\,,
\end{align}
where $\vec P = \vec p_1 + \vec p_2$ is the total momentum. The corresponding second-order
contribution is thus of the form
\begin{align}
\frac{2\alpha}{\pi }\bigg(\mathcal{I}_\epsilon-\frac{1}{6\epsilon}\bigg)
\,\langle H'^{(4)}\,\frac{1}{(E_0-H_0)'}\,\big[\vec P,\big[V,\vec P\big]\big]\rangle \,.
\end{align}
In order to identify divergencies in this second-order matrix element, we introduce the following
identities
\begin{eqnarray}\label{trans:1}
H'^{(4)} &=& \{H_0-E_0,Q\}+H_R\,,\\ \label{trans:2}
{} [P^i,[V,P^i]]&=& \{H_0-E_0,Q'\}+H_R'\,,
\end{eqnarray}
where the operators $H_R$ and $H_R'$ are chosen in such a way that the second-order
corrections with these operators are finite. We have
\begin{eqnarray}
Q &=& -\frac14\bigg[\frac{Z\alpha}{r_1}+\frac{Z\alpha}{r_2}\bigg]_\epsilon+\frac{\alpha}{2r}\,,\\
Q'&=& 2\bigg[\frac{Z\alpha}{r_1}+\frac{Z\alpha}{r_2}\bigg]_\epsilon\,,
\end{eqnarray}
and the regularized operators $H_R$ and $H_R'$ are defined by their action on ket states as
\begin{align}
H_R|\phi\rangle =&\ \bigg[-\frac12(E_0-V)^2-\frac{Z\alpha}{4}\frac{\vec r_1\cdot\vec\nabla_1}{r_1^3}
-\frac{Z\alpha}{4}\frac{\vec r_2\cdot\vec\nabla_2}{r_2^3}
 \nonumber \\ &
+\frac14\nabla_1^2\nabla_2^2
- p_1^i\frac{\alpha}{2r}\bigg(\delta^{ij}+\frac{r^i r^j}{r^2}\bigg)\,p_2^j\bigg]|\phi\rangle\,,
 \\
H_R'|\phi\rangle =&\ -2Z\alpha\bigg(\frac{\vec r_1\cdot\vec\nabla_1}{r_1^3}
+\frac{\vec r_2\cdot\vec\nabla_2}{r_2^3}\bigg)|\phi\rangle
\,.
\end{align}
\begin{widetext}
We now rewrite the second-order correction as
\begin{eqnarray}\label{sec:1}
\bigg\langle[P^i,[V,P^i]]\frac{1}{(E_0-H_0)'}H'^{(4)}\bigg\rangle
&=&\bigg\langle[P^i,[V,P^i]]\frac{1}{(E_0-H_0)'}H_R\bigg\rangle
+\langle \big(\langle Q\rangle-Q)[P^i,[V,P^i]]\rangle
\\
&=&\bigg\langle H_R'\frac{1}{(E_0-H_0)'}H_R\bigg\rangle
+\langle H_R\,\big(\langle Q'\rangle-Q'\big)\rangle
+\langle \big(\langle Q\rangle-Q\big)\,[P^i,[V,P^i]]\rangle
\label{eq:113}
\,.
\end{eqnarray}
The first term in the last equation
is finite. The third term is evaluated as
\begin{eqnarray}
\langle \big(\langle Q\rangle-Q\big)\,[P^i,[V,P^i]]\rangle
&=&\bigg\langle\bigg(\frac{E_0}{2}+\bigg\langle\frac{\alpha}{4r}\bigg\rangle
+\frac{(Z-2)\alpha}{4r_2}\bigg)\,4\pi Z\alpha\,\delta^3(r_1) + (1\leftrightarrow2)\bigg\rangle\,.
\end{eqnarray}
Here we used the identity $\delta^d(r)\,[1/r]_\epsilon = 0$, which is valid in dimensional regularization.
The evaluation of the second term in Eq.~(\ref{eq:113}) is more complicated. We rewrite it as
\begin{eqnarray}
\langle H_R\,\big(\langle Q'\rangle-Q'\big)\rangle
&=& E^{(4)}\langle Q'\rangle - \langle \big(H'^{(4)}-\{H_0-E_0,Q\}\big)\,Q'\rangle
\rangle\nonumber\\
&=&\bigg\langle E^{(4)}\bigg(\frac{2\alpha}{r}-4E_0\bigg)+\frac12\big[(\vec\nabla_1 Q)\cdot(\vec\nabla_1 Q')+
(\vec\nabla_2 Q)\cdot(\vec\nabla_2 Q')\big] - H'^{(4)}\,Q'\bigg\rangle
= X_1 + X_2 + X_3\,,
\end{eqnarray}
where $E^{(4)} = \langle H^{(4)}\rangle = \langle H_R\rangle$ since $\langle H''^{(4)}\rangle=0$.
Term $X_1$ needs no further simplification,
\begin{equation}
X_1 = \bigg\langle E^{(4)}\bigg(\frac{2\alpha}{r}-4E_0\bigg)\bigg\rangle
\,.
\end{equation}
Term $X_2$ reduces to
\begin{eqnarray}
X_2=\frac12\big[(\vec\nabla_1 Q)\cdot(\vec\nabla_1 Q')+
(\vec\nabla_2 Q)\cdot(\vec\nabla_2 Q')\big]
&=&-\frac{1}4\bigg[\frac{(Z\alpha)^2}{r_1^4}+\frac{(Z\alpha)^2}{r_2^4}\bigg]_\epsilon + \frac12\bigg(\frac{Z\alpha\,\vec r_1}{r_1^3}
-\frac{Z\alpha\,\vec r_2}{r_2^3}\bigg)\cdot\frac{\alpha\,\vec r}{r^3}\,,
\end{eqnarray}
where $[(Z\,\alpha)^2/r_1^4]_\epsilon = [\nabla (Z\,\alpha/r_1)]^2_\epsilon$. Term $X_3$  for triplet states is
\begin{eqnarray}
X_3&=&\langle -H'^{(4)}\,Q'\rangle =\bigg\langle\bigg[\frac18(p_1^4+p_2^4)-\frac{Z\pi\alpha}{2}\big[\delta^3(r_1)
+\delta^3(r_2)\big]+\frac{\alpha}{2} p_1^i\bigg(\frac{\delta^{ij}}{r}+\frac{r^i r^j}{r^3}\bigg)\,p_2^j\bigg]\,Q'\bigg\rangle\nonumber\\
&=& \bigg\langle\,\frac18(p_1^4+p_2^4)\,Q'-(Z\alpha)^2\pi\bigg[\frac{\delta^3(r_1)}{r_2}
+\frac{\delta^3(r_2)}{r_1}\bigg]+p_1^i\bigg(\frac{Z\alpha}{r_1}+\frac{Z\alpha}{r_2}\bigg)
\frac{\alpha}{r}\bigg(\delta^{ij}+\frac{r^i r^j}{q^2}\bigg)\,p_2^j \bigg\rangle
\,.
\end{eqnarray}
The first term in the last equation contains singularities. We transform it as
\begin{eqnarray}
\frac18(p_1^4+p_2^4)\,Q' &=& \frac18\big[(p_1^2+p_2^2)^2-2\,p_1^2\,p_2^2\big]\,Q'
=\frac18\bigg[(p_1^2+p_2^2)\,Q'\,(p_1^2+p_2^2)+\frac12\big[p_1^2+p_2^2,\big[p_1^2+p_2^2,Q'\big]\big]
-2\,p_1^2\,Q'\,p_2^2\bigg]
\nonumber\\
&=&\frac12\bigg[2(E_0-V)^2\bigg[\frac{Z\alpha}{r_1}+\frac{Z\alpha}{r_2}\bigg]_\epsilon
+\bigg\{\bigg[\frac{(Z\alpha)^2}{r^4_1}\bigg]_\epsilon
-\frac{Z\alpha\,\vec r_1}{r_1^3}\cdot\frac{\alpha\,\vec r}{r^3}
+(1\leftrightarrow2)\bigg\}
-\,p_1^2\bigg(\frac{Z\alpha}{r_1}+\frac{Z\alpha}{r_2}\bigg)\,p_2^2\bigg]
\,.\nonumber\\
\end{eqnarray}
It is convenient to convert $[Z^3/r_a^3]_\epsilon$ into $[Z^2/r_a^4]_\epsilon$, which is
achieved by the following identity
\begin{eqnarray}
\left\langle\bigg[\frac{(Z\alpha)^2}{r_1^4}\bigg]_\epsilon\right\rangle
&=&\left\langle-2 \bigg[\frac{Z\alpha}{r_1}\bigg]^3_\epsilon-2\,Y_1\right\rangle
\,,
\end{eqnarray}
where
\begin{equation}
Y_1 = \bigg(E_0+\frac{Z\alpha}{r_2}-\frac{\alpha}{r}-\frac{p_2^2}{2}\bigg)\frac{(Z\alpha)^2}{r_1^2}-\frac12\,\vec p_1\frac{(Z\alpha)^2}{r_1^2}\,\vec p_1\,.
\end{equation}
The final result for the second-order contribution is a sum $X_1+X_2+X_3$ and takes the form
\begin{eqnarray}
E^{(7)}_{\rm sec, SE}\label{Eseco:tot}
&=&
2\,\langle H''^{(4)}\,\frac{1}{(E_0-H_0)'}\,H''^{(5)}_{\rm SE}\rangle
+\frac{\alpha}{\pi}\bigg[\frac59+\frac23\ln(\alpha^{-2})-\frac23\ln(2\lambda)\bigg]
 \bigg(
\bigg\langle H_R'\frac{1}{(E_0-H_0)'}H_R\bigg\rangle
  \nonumber \\ &&
+\bigg\langle
E^{(4)}\bigg(\bigg\langle\frac{2\alpha}{r}\bigg\rangle-4 E_0\bigg)
+\bigg\{
\bigg(2E_0+\bigg\langle\frac{\alpha}{r}\bigg\rangle\bigg)\,\pi Z \alpha\,\delta^3(r_1)
+\bigg(E_0+\frac{Z\alpha}{r_2}-\frac{\alpha}{r}\bigg)^2\,\frac{Z\alpha}{r_1}
+2\bigg(E_0+\frac{Z\alpha}{r_2}-\frac{\alpha}{r}\bigg)\,\frac{(Z\alpha)^2}{r^2_1}\nonumber\\
&&-Y_1
-\frac12p_1^2\,\frac{Z\alpha}{r_1}\,p_2^2
-\frac{1}{4} \bigg[\frac{(Z\alpha)^2}{r_1^4}\bigg]_\epsilon
- 2\pi Z\alpha^2
\frac{\delta^3(r_1)}{r_2}
+p_1^i\,\frac{Z\alpha}{r_1}\frac{\alpha}{r}\bigg(\delta^{ij}+\frac{r^i r^j}{r^2}\bigg)\,p_2^j
+(1\leftrightarrow2)\bigg\}\bigg\rangle
\bigg)
\,.
\nonumber \\
\end{eqnarray}
Note that the above expression is written completely in the coordinate representation. In the
following derivation of other contributions, we will often find it convenient to keep the
two-body electron-electron terms in the momentum representation. This is advantageous because in
the momentum representation the evaluation of the electron-electron terms is simpler.

\end{widetext}

\section{Effective $\bm{\alpha^7m}$ Hamiltonian}
\label{sec:hamma7}

We now turn to the derivation of the first-order self-energy $\alpha^7m$  contribution
represented by the first term in Eq.~(\ref{eq:1}). This correction  can be conveniently split
into three parts, according to the region of the contributing photon momenta,
\begin{equation} \label{eq:2}
\big< H^{(7)}_{\rm SE}\big> = E_L^{\Lambda}+E_M+E_H\,.
\end{equation}
Here, the three contributions $E_L^{\Lambda}$, $E_M$, and $E_H$ are induced by momenta $k$ of the
radiative photon of the order $m\alpha^2$, $m\alpha$, and $m$, respectively. These three
terms will be referred to as the low-energy, middle-energy, and high-energy parts,
correspondingly.

\subsection{Low-energy part}

We now turn to the evaluation of relativistic corrections of order $\alpha^7m$ to the
leading-order one-loop nonrelativistic dipole self-energy contribution, $E_{L0}$, given by
Eq.~(\ref{eq:L:0}). Contributions arising from the small-$k$ region, $k < \Lambda$, give rise to
the relativistic corrections to the Bethe logarithm, already computed in
Ref.~\cite{yerokhin:18:betherel}. So, in the present work we are only concerned with the
large-$k$ region, $k > \Lambda$.

The $\alpha^7m$ corrections to $E_{L0}^{\Lambda}$ arise as: ({\em i}) perturbations of the
reference-state wave function $\phi$, the zeroth-order energy $E_0$ and the zeroth-order
Hamiltonian $H_0$ by the Breit Hamiltonian $H^{(4)}$, ({\em ii}) a perturbation of the current
$\vec{p} \to \delta \vec{j}$, and ({\em iii}) a retardation (quadrupole) correction. The
corresponding $\alpha^7m$  corrections will be denoted as $E_{L1}$, $E_{L2}$, and $E_{L3}$,
respectively,
\begin{align}\label{eq:low:1}
E_L^\Lambda = E_{L1}^{\Lambda} + E_{L2}^{\Lambda} + E_{L3}^{\Lambda}\,.
\end{align}
Our calculation of these terms will be similar to that of the low-energy photon-exchange
contributions, described in Sec.~III of Ref.~\cite{patkos:20}.

\subsubsection{$E_{L1}^\Lambda$}

The first term $E_{L1}^\Lambda$ is due to a perturbation by the Breit Hamiltonian and is written
as
\begin{eqnarray}\label{eq:L:1}
E_{L1}^{\Lambda} &=& e^2\int_{\Lambda}^\infty \frac{d^dk}{(2\pi)^d2k}\,
 \delta_\perp^{ij}(k)
\,\delta\bigg\langle\phi\bigg|p_1^i\frac{1}{E_0-H_0-k}p_1	^j\bigg|\phi\bigg\rangle
 \nonumber \\ &&
+(1\leftrightarrow2)\,.
\end{eqnarray}
Here the symbol $\delta\langle\cdots\rangle$ denotes the first-order perturbation of the matrix
element $\langle\cdots\rangle$ by the Breit Hamiltonian $H^{(4)}$, which implies perturbations of
the reference-state wave function $\phi$, the energy $E_0$, and the zeroth-order Hamiltonian
$H_0$. Since $k$ is much bigger than $H_0-E_0$, we can expand the integrand of Eq.~(\ref{eq:L:1})
in large $k$, keeping only the $1/k^2$ term, while $1/k$ contributes at the lower order of
$\alpha^6\,m$. The result is
\begin{eqnarray}\label{eq:L:2}
 &&  \!\!\!\!\!\!\!\!\!\!
 \delta\Big< \phi\Big|  p_1^i\frac{1}{E_0-H_0-k}p_1^j\Big|\phi\Big> + (1\leftrightarrow2)
\nonumber\\
&=&\frac{1}{2\,k^2}\,\delta\Big<\phi\Big|\big[p_1^i,\big[H_0-E_0,p_1^j\big]\big]\Big|\phi\Big>
+(1\leftrightarrow2)\nonumber\\ \nonumber\\
&=& \frac{1}{k^2}\,\Big<\phi\Big|[p_1^i,[V,p_1^j]]\frac{1}{(E_0-H_0)'}H^{(4)}\Big|\phi\Big>
\nonumber\\&&
+ \frac{1}{2\,k^2}\,\Big<\phi\Big|[p_1^i,[H^{(4)},p_1^j]]\Big|\phi\Big>
+(1\leftrightarrow2)\,.
\end{eqnarray}
The first term in the last equation is the second-order contribution already accounted for in the
previous section, and thus it will be omitted here. Note that the above expansion is valid up to the
electron-nucleus Dirac $\delta$-function terms, which we are omitting for the present. They will
be restored  later by matching our calculation against the hydrogenic result. The same will apply
also for $E^\Lambda_{L2}$ and $E^\Lambda_{L3}$.

After expanding in $\epsilon = (3-d)/2$ and then in $\alpha$, we obtain the result for the
corresponding effective Hamiltonian $H_{L1}$, $E_{L1}^\Lambda = \langle H_{L1}\rangle$,
\begin{widetext}
\begin{eqnarray}
H_{L1} &=& \alpha^2\bigg(\frac{5}{36}+\frac{1}{12}\ln\Lambda_{\epsilon}\bigg)
Z\,\big[\nabla_1^2\delta^{d}(r_1)+\nabla_2^2\delta^{d}(r_2)\big]
-\alpha^2\bigg\{\sigma_1\cdot\sigma_2\bigg(-\frac{7}{27}-\frac19\ln{\Lambda_{\epsilon}}
\bigg)\,q^2
\nonumber\\&&
+\bigg(-\frac{5}{9}-\frac13\ln{\Lambda_{\epsilon}}\bigg)\,
\bigg[q^2+4\vec P_1\cdot\vec P_2 - 4\frac{\big(\vec P_1\cdot \vec q\big)\big(\vec P_2\cdot\vec q\big)}{q^2}\bigg]\bigg\}
\end{eqnarray}
where $\vec P_1 = \frac12(\vec p_1+\vec p_1{}')$ and $\vec P_2 = \frac12(\vec p_2+\vec p_2{}')$
are sums of the {\em in} and {\em out} momenta of the corresponding electron, and $\vec q = \vec p_1{}' - \vec p_1$. Using the identity
valid for the expectation values of the operators,
\begin{align}\label{14}
\nabla_1^2\delta^{d}(r_1) =
-2\,p_1^2\,\delta^{d}(r_1)+2\,\vec p_1\,\delta^{d}(r_1)\,\vec p_1
=-4\,(E_0-V-p_2^2/2)\,\delta^{d}(r_1)+2\,\vec p_1\,\delta^{d}(r_1)\,\vec p_1\,,
\end{align}
we simplify the expression further, obtaining the final result
\begin{eqnarray}
H_{L1}&=& \alpha^2
\bigg\{Z \delta^{d}(r_1)
\bigg(-\frac59-\frac13\ln\Lambda_{\epsilon}\bigg)\bigg(E_0-V-\frac{p_2^2}{2}\bigg)
+\bigg(\frac{5}{18}+\frac{1}{6}\ln\Lambda_{\epsilon}\bigg)\,\vec p_1\,Z\delta^{d}(r_1)\,\vec p_1
+(1\leftrightarrow2)\bigg\}\nonumber\\
&&+\alpha^2\bigg\{\sigma_1\cdot\sigma_2\bigg(\frac{7}{27}+\frac19\ln{\Lambda_{\epsilon}}
\bigg)\,q^2
+\bigg(\frac{5}{9}+\frac13\ln{\Lambda_{\epsilon}}\bigg)\,
\bigg[q^2+4\vec P_1\cdot\vec P_2 - 4\frac{\big(\vec P_1\cdot \vec q\big)\big(\vec P_2\cdot\vec q\big)}{q^2}\bigg]\bigg\}\,.
\end{eqnarray}
Note that in the above expression we keep the electron-nucleus terms in the coordinate
representation but the electron-electron terms in the momentum representation.

\subsubsection{$E_{L2}^\Lambda$}
The second term in Eq.~(\ref{eq:low:1}), $E_{L2}^\Lambda$, comes from a correction to the
current. Specifically, $\vec{p}_1$ gets a correction $\delta \vec j_1$, which is
\begin{eqnarray}
\delta j^i_1 = {\rm i}\big[H^{(4)},r_1^i\big]
=-\frac{1}{2} p_1^i p_1^2 - \frac{\alpha}{2}\bigg[\frac{\delta^{ij}}{r}+\frac{r^i r^j}{r^3}\bigg]_\epsilon\,p_2^j\,,
\end{eqnarray}
and the same for $\vec{p}_2$. The contribution
$E_{L2}^{\Lambda}$ is then
\begin{eqnarray}
E_{L2}^{\Lambda} &=& 2e^2\int_\Lambda^\infty \frac{d^dk}{(2\pi)^d2k}\,\delta_\perp^{ij}(k)
\,\bigg\langle\phi\bigg|\delta j_1^i\frac{1}{E_0-H_0-k}p_1^j\bigg|\phi\bigg\rangle
+(1\leftrightarrow2)\,.
\end{eqnarray}
Expanding this expression in large $k$ and performing the angular average, we arrive at
\begin{eqnarray}
E_{L2}^{\Lambda} = e^2\frac{d-1}{d}\int_\Lambda^\infty\frac{d^dk}{(2\pi)^d2k^3}
 \big\langle\phi\big|\,[\delta j_1^i,[V,p_1^i]]\,\big|\phi\big\rangle
+(1\leftrightarrow2)= E_{L2}^A+E_{L2}^B+E_{L2}^C\,.
\end{eqnarray}
This expression consists of three-body and two-body terms.
The three-body contribution $E_{L2}^A$ is due to
the first term in $\delta \vec j_1$; it is transformed as
\begin{eqnarray}
E_{L2}^A&=&-e^2\frac{d-1}{2d}\int_\Lambda^\infty\frac{d^dk}{(2\pi)^d2k^3}
\bigg\langle\phi\bigg|\bigg[p_1^i\,p_1^2,\bigg[-\bigg[\frac{Z\alpha}{r_1}\bigg]_\epsilon,p_1^i\bigg]\bigg]\bigg|\phi\bigg\rangle
+(1\leftrightarrow2)\nonumber\\
&=& \frac{\alpha}{\pi}\bigg(-\frac{5}{18}-\frac16\ln\Lambda_{\epsilon}\bigg)
\bigg\langle\phi\bigg|-\bigg[V,\bigg[p_1^2,\bigg[\frac{Z\alpha}{r_1}\bigg]_\epsilon\bigg]\bigg]
+8\pi Z\alpha\,\delta^{d}(r_1)\,\big(E_0-V\big)
-\vec p_2\,4\pi Z\alpha\,\delta^{d}(r_1)\,\vec p_2\bigg|\phi\bigg\rangle
+(1\leftrightarrow2)\,.\nonumber\\
\end{eqnarray}
There are two two-body contributions, the first one coming from the term $[-\frac12
p_1^i\,p_1^2,[\alpha/r,p_1^i]]$. We evaluate it by switching into the momentum representation,
\begin{eqnarray}
E_{L2}^B&=&e^2\frac{d-1}{d}\int_\Lambda^\infty\frac{d^dk}{(2\pi)^d2k^3}
\bigg\langle\phi\bigg|
\bigg[-\frac12 p_1^i\,p_1^2,\bigg[\bigg[\frac{\alpha}{r}\bigg]_\epsilon,p_1^i\bigg]\bigg]\bigg|\phi\bigg\rangle
+(1\leftrightarrow2)\nonumber\\
&=&\alpha^2\bigg(\frac{20}{9}+\frac43\ln{\Lambda_{\epsilon}}\bigg)\bigg\langle\phi\bigg|
\frac12\big(\vec P_1-\vec P_2\big)^2+\vec P_1\cdot\vec P_2+\frac14 q^2
+\frac{\big[\big(\vec P_1-\vec P_2\big)\cdot\vec q\big]^2}{q^2}
+2\frac{\big(\vec P_1\cdot\vec q\big)\big(\vec P_2\cdot\vec q\big)}{q^2}\bigg|\phi\bigg\rangle
\,.
\end{eqnarray}
The two-body contribution induced by the second term in $\delta \vec j_1$ is evaluated with help of
integration formulas from Appendix~C of Ref.~\cite{patkos:20} as
\begin{eqnarray}
E_{L2}^C&=&e^2\frac{d-1}{d}\int_\Lambda^\infty\frac{d^dk}{(2\pi)^d2k^3}
 \big\langle\phi\big|\,\bigg(- \frac{\alpha}{2}\bigg[\frac{\delta^{ij}}{r}+\frac{r^i r^j}{r^3}\bigg]_\epsilon\bigg)
\bigg[\,p_2^j,\bigg[\bigg[\frac{\alpha}{r}\bigg]_\epsilon,p_1^i\bigg]\bigg]\,\big|\phi\big\rangle
+(1\leftrightarrow2)\nonumber\\
&=&
\pi \alpha^3\bigg\langle\phi\bigg|\bigg(-\frac{4}{9}-\frac23\ln{\Lambda_{\epsilon}}-\frac43\ln2+\frac43\ln q\bigg)q\bigg|\phi\bigg\rangle
\,.
\end{eqnarray}

Adding together the individual contributions to $E_{L2}^\Lambda$ and using the expectation value
identity
\begin{equation}\label{2ph:to:3ph}
\alpha\frac{\big[(\vec P_1-\vec P_2)\cdot\vec q\big]^2}{q^2}
=\bigg\{\frac{1}{4\pi}\bigg[\frac{Z\alpha\,\vec r_1}{r_1^3}\bigg]_\epsilon\cdot\bigg[\frac{\alpha\,\vec r}{r^3}\bigg]_\epsilon
+(1\leftrightarrow2)\bigg\}
+\frac{\pi\alpha^2}{2}\big\{1+\epsilon\,(2\ln2-2\ln q)\big\}q\,,
\end{equation}
we write the result in terms of the effective Hamiltonian $H_{L2}$, $E_{L2}^\Lambda=\langle
H_{L2}\rangle$, which is
\begin{align}
H_{L2}=&\ \frac{\alpha}{\pi}\bigg\{\bigg(-\frac{5}{9}-\frac13\ln\Lambda_{\epsilon}\bigg)
\bigg(\bigg[\frac{ (Z\alpha)^2}{r_1^4}\bigg]_\epsilon
-2\bigg[\frac{Z\alpha\,\vec r_1}{r_1^3}\bigg]_\epsilon\cdot\bigg[\frac{\alpha\,\vec r}{r^3}\bigg]_\epsilon
+4\pi Z\alpha\,\delta^{d}(r_1)\bigg(E_0-V-\frac{p_2^2}{2m}\bigg)\bigg)
+(1\leftrightarrow2)\bigg\}\nonumber\\
&\ + \alpha^2\bigg(\frac{20}{9}+\frac43\ln{\Lambda_{\epsilon}}\bigg)
\bigg(\frac12\big(\vec P_1-\vec P_2\big)^2+\vec P_1\cdot\vec P_2+\frac14 q^2
+2\frac{\big(\vec P_1\cdot\vec q\big)\big(\vec P_2\cdot\vec q\big)}{q^2}\bigg)
+\frac{2\pi}{3} \alpha^3 \,q\,.
\end{align}

\subsubsection{$E_{L3}^\Lambda$}

The third term in Eq.~(\ref{eq:low:1}), $E_{L3}^\Lambda$, is a retardation correction. It can be
expressed as
\begin{eqnarray}
E_{L3}^{\Lambda} &=& e^2\int_\Lambda^\infty \frac{d^dk}{(2\pi)^d2k}
\, \delta_\perp^{ij}(k)
\,\delta_{k^2} \bigg\langle\phi\bigg|p_1^i\,e^{{\rm i}\vec k\cdot\vec r_1}\frac{1}{E_0-H_0-k}p_1^j\,e^{-{\rm i}\vec k\cdot\vec r_1}\bigg|\phi\bigg\rangle
+(1\leftrightarrow2)\,,
\end{eqnarray}
where the symbol $\delta_{k^2}\big<\ldots\big>$ means that the exponential factors $e^{{\rm
i}\vec k\cdot\vec r_1}$ and $e^{-{\rm i}\vec k\cdot\vec r_1}$ in the matrix element
$\big<\ldots\big>$ are expanded in small $k$ up to order $k^2$. Because $\Lambda$ is arbitrarily
large, we perform the large-$k$ expansion of the resolvent,
\begin{eqnarray}
\frac{1}{E_0-H_0-k} &=& -\frac{1}{k}+\frac{H_0-E_0}{k^2}-\frac{(H_0-E_0)^2}{k^3}+\frac{(H_0-E_0)^3}{k^4}+\cdots\,.
\end{eqnarray}
The $k$ expansion needs to be extended up to order $k^{-4}$ because of the additional $k^2$ from
the expansion of the exponential factors. The resulting correction of order $\alpha^7\,m$ is
\begin{eqnarray}\label{eq:201}
E_{L3}^{\Lambda} &=& e^2\int_\Lambda^\infty \frac{d^dk}{(2\pi)^d2k^5}\,\delta_\perp^{ij}(k)
\,\delta_{k^2} \bigg\langle\phi\bigg|p_1^i\,e^{{\rm i}\vec k\cdot\vec r_1}(H_0-E_0)^3\,p_1^j
\,e^{-{\rm i}\vec k\cdot\vec r_1}\bigg|\phi\bigg\rangle
+(1\leftrightarrow2)\,.
\end{eqnarray}
The radial integration $k$ can be performed in the same way as in the previous low-energy contributions.
For the angular integration we will use the formulas from Appendix~C of Ref.~\cite{patkos:20}.
The matrix element in Eq.~(\ref{eq:201}) can be simplified by using identity $ \exp({\rm i}\vec
k\cdot\vec r)\,f(\vec p)\,\exp(-{\rm i}\vec k\cdot\vec r) = f(\vec p-\vec k)\,$. We, therefore,
have
\begin{eqnarray}
\delta_{k^2}\bigg[p_1^i\,e^{{\rm i}\vec k\cdot\vec r_1}(H_0-E_0)^3\,p_1^j
\,e^{-{\rm i}\vec k\cdot\vec r_1}\bigg]
&=& \delta_{k^2}\bigg[p_1^i\,\bigg(H_0-E_0-\vec p_1\cdot\vec k+\frac{k^2}{2}\bigg)^3\,p_1^j\bigg]\nonumber\\
&=&p_1^i\bigg[\frac32(H_0-E_0)^2\,k^2+\vec p_1\cdot\vec k\,(H_0-E_0)\,\vec p_1\cdot\vec k
+2\,\big(\vec p_1\cdot\vec k\big)^2\,(H_0-E_0)\bigg]\,p_1^j\,.
\end{eqnarray}
We will refer to the contributions induced by the three terms in the brackets of the above
expression as $E_{L3}^A$, $E_{L3}^B$, and $E_{L3}^C$, respectively,
\begin{equation}
E_{L3}^\Lambda = E_{L3}^A+E_{L3}^B+E_{L3}^C\,.
\end{equation}
Starting with the three-photon contribution $E_{L3}^A$, we obtain
\begin{eqnarray}
E_{L3}^A &=& \frac{3e^2}{2}\int_\Lambda^\infty \frac{d^dk}{(2\pi)^d2k^3}\,\delta_\perp^{ij}(k)
\,\bigg\langle\phi\bigg|p_1^i\,(H_0-E_0)^2\,p_1^j\bigg|\phi\bigg\rangle
+(1\leftrightarrow2)\,.
\end{eqnarray}
Averaging this expression over the angular variables, commuting $H_0-E_0$ to the left and to the
right, and performing the integration, we obtain
\begin{eqnarray}
E_{L3}^A &=&\bigg\langle\phi\bigg|\frac{\alpha}{\pi}\bigg\{\bigg(\frac{5}{6}+\frac12\ln\Lambda_{\epsilon}\bigg)
\bigg[\frac{(Z\alpha)^2}{r_1^4}\bigg]_\epsilon
-\bigg(\frac{5}{3}+\ln\Lambda_{\epsilon}\bigg)
\bigg[\frac{Z\alpha\,\vec r_1}{r_1^3}\bigg]_\epsilon\cdot\bigg[\frac{\alpha\,\vec r}{r^3}\bigg]_\epsilon
+(1\leftrightarrow2)\bigg\}
  \nonumber\\&&
+\pi\alpha^3\bigg(-\frac{5}{3}-\ln\Lambda_{\epsilon}-2\ln2+2\ln q\bigg)\,q\bigg|\phi\bigg\rangle\,.
\end{eqnarray}
The contribution of the second term is evaluated as
\begin{eqnarray}
E_{L3}^B &=& e^2\int_\Lambda^\infty \frac{d^dk}{(2\pi)^d2k^5}\,\delta_\perp^{ij}(k)k^m k^n
\,\bigg\langle\phi\bigg|p_1^i p_1^m\,(H_0-E_0)\,p_1^n p_1^j\bigg|\phi\bigg\rangle
+(1\leftrightarrow2)\nonumber\\
&=&\frac{e^2}{d(d+2)}\int_\Lambda^\infty \frac{d^dk}{(2\pi)^d2k^3}
\,\bigg\langle\phi\bigg|d\,p_1^i p_1^j\,(H_0-E_0)\,p_1^i p_1^j
-p_1^2\,(H_0-E_0)\,p_1^2\bigg|\phi\bigg\rangle
+(1\leftrightarrow2)
\,.
\end{eqnarray}
We transform this expression further with the help of the following identities,
\begin{eqnarray}
p_1^2\,(H_0-E_0)\,p_1^2 &=& \frac12\,\big[p_1^2,\big[H_0-E_0,p_1^2\big]\big]
= \frac12\,\big[p_1^2,\big[V,p_1^2\big]\big]
\,,\\
p_1^i\,\big[p_1^i,\big[V,p_1^j\big]\big]\,p_1^j &=&
\frac14\,\big[p_1^2,\big[V,p_1^2\big]\big] - \frac12\,p_1^2\,\big[\,p_1^i,\big[V,p_1^i\big]\big]
+\frac12\,p_1^i\,\big[p_1^j,\big[V,p_1^j\big]\big]\,p_1^i
\,,
\end{eqnarray}
and
\begin{eqnarray}
p_1^i p_1^j\,(H_0-E_0)\,p_1^i p_1^j
&=& \frac12\,p_1^2\,\big[\,p_1^i,\big[V,p_1^i\big]\big]+\frac14\,\big[p_1^2,\big[V,p_1^2\big]\big]
+\frac12\,p_1^i\,\big[p_1^j,\big[V,p_1^j\big]\big]\,p_1^i
\,.
\end{eqnarray}
The result for the second term is
\begin{eqnarray}
E_{L3}^B
&=& \frac{\alpha}{\pi}\,\bigg\langle\phi\bigg|\bigg(\frac{2}{225}+\frac{1}{120}\ln\Lambda_{\epsilon}\bigg)
\big[p_1^2,\big[V,p_1^2\big]\big]
+\bigg(\frac{3}{25}+\frac{1}{20}\ln\Lambda_{\epsilon}\bigg)
\nonumber\\&&\times
\bigg(p_1^i\,\big[p_1^j,\big[V,p_1^j\big]\big]\,p_1^i
+p_1^2\,\big[\,p_1^i,\big[V,p_1^i\big]\big] \bigg)\bigg|\phi\bigg\rangle+(1\leftrightarrow2)\nonumber\\
&=& \bigg\langle\phi\bigg|\frac{\alpha}{\pi}\,\bigg\{
\bigg(\frac{8}{225}+\frac{1}{30}\ln\Lambda_{\epsilon}\bigg)\bigg(\bigg[\frac{(Z\alpha)^2}{r_1^4}\bigg]_\epsilon-
\bigg[\frac{Z\alpha\,\vec r_1}{r_1^3}\bigg]_\epsilon\cdot\bigg[\frac{\alpha\,\vec r}{r^3}\bigg]_\epsilon\bigg)
+\bigg(\frac{3}{25}+\frac{1}{20}\ln\Lambda_{\epsilon}\bigg)\nonumber\\
&&
\times\bigg(p_1^i\,4\pi Z\alpha\,\delta^{d}(r_1)\,p_1^i
+\bigg(E_0-V-\frac{p_2^2}{2}\bigg)\,8\pi Z\alpha\,\delta^{d}(r_1) \bigg)
+(1\leftrightarrow2)\bigg\}\nonumber\\
&&+\alpha^2\bigg\{-\frac{32}{225}\frac{\big[(\vec P_1-\vec P_2)\cdot\vec q\big]^2}{q^2}
-\frac{24}{25}(\vec P_1-\vec P_2)^2-\frac{48}{25}\vec P_1\cdot\vec P_2
-\frac{64}{225}\frac{\big(\vec P_1\cdot\vec q\big)\big(\vec P_2\cdot\vec q\big)}{q^2}\nonumber\\
&&+\ln\Lambda_{\epsilon}\bigg(-\frac25(\vec P_1-\vec P_2)^2
-\frac{2}{15}\frac{\big[(\vec P_1-\vec P_2)\cdot\vec q\big]^2}{q^2}
-\frac45 \vec P_1\cdot\vec P_2-\frac{4}{15}\frac{\big(\vec P_1\cdot\vec q\big)\big(\vec P_2\cdot\vec q\big)}{q^2}\bigg)\bigg\}
\bigg|\phi\bigg\rangle
\,.
\end{eqnarray}
Finally, the term $E_{L3}^C$ is calculated as
\begin{eqnarray}
E_{L3}^C &=& 2e^2\int_\Lambda^\infty \frac{d^dk}{(2\pi)^d2k^5}\,\delta_\perp^{ij}(k)k^m k^n
\,\bigg\langle\phi\bigg|p_1^i p_1^m p_1^n\,(H_0-E_0)\,p_1^j\bigg|\phi\bigg\rangle
+(1\leftrightarrow2)\nonumber\\
&=& e^2\int_\Lambda^\infty \frac{d^dk}{(2\pi)^d2k^3}\bigg\langle\phi\bigg|
\frac{(d-1)}{d(d+2)}
\,\big[p_1^i p_1^2,\big[V,p_1^i\big]\big]\bigg|\phi\bigg\rangle
+(1\leftrightarrow2)\nonumber\\
&=&\bigg\langle\phi\bigg|\frac{\alpha}{\pi}\bigg\{\bigg(\frac{62}{225}+\frac{2}{15}\ln\Lambda_{\epsilon}\bigg)
\bigg(4\pi Z\alpha\,\delta^{d}(r_1)\bigg(E_0-V-\frac{p_2^2}{2}\bigg)
+\bigg[\frac{(Z\alpha)^2}{r_1^4}\bigg]_\epsilon
-\bigg[\frac{Z\alpha\,\vec r_1}{r_1^3}\bigg]_\epsilon\cdot\bigg[\frac{\alpha\,\vec r}{r^3}\bigg]_\epsilon\bigg)
+(1\leftrightarrow2)\bigg\}\nonumber\\
&&
+\alpha^2\bigg(\frac{124}{225}+\frac{4}{15}\ln\Lambda_{\epsilon}\bigg)
\bigg(-\big(\vec P_1-\vec P_2\big)^2-\frac12 q^2
-2\vec P_1\cdot\vec P_1
-2\frac{\big[(\vec P_1-\vec P_2)\cdot\vec q\big]^2}{q^2}
-4\frac{\big(\vec P_1\cdot\vec q\big)\big(\vec P_2\cdot\vec q\big)}{q^2}\bigg)\bigg|\phi\bigg\rangle
\,.
\end{eqnarray}
Adding together $E_{L3}^A$, $E_{L3}^B$, and $E_{L3}^C$ and transforming the sum with help of
Eq.~(\ref{2ph:to:3ph}), we obtain the retardation correction $E_{L3}^\Lambda$ as
\begin{eqnarray}
E_{L3}^\Lambda
&=&\bigg\langle\phi\bigg|\frac{\alpha}{\pi}
\bigg\{\frac{464}{225}\pi Z\alpha\,\delta^{d}(r_1)\bigg(E_0-V-\frac{p_2^2}{2}\bigg)
+\frac{103}{90}
\bigg[\frac{(Z\alpha)^2}{r_1^4}\bigg]_\epsilon
-\frac{103}{45}
\bigg[\frac{Z\alpha\,\vec r_1}{r_1^3}\bigg]_\epsilon\cdot\bigg[\frac{\alpha\,\vec r}{r^3}\bigg]_\epsilon\nonumber\\
&&
+\frac{3}{25}p_1^i\,4\pi Z\alpha\,\delta^{d}(r_1)\,p_1^i
+\ln\Lambda_{\epsilon}\bigg(
\frac{14}{15}\pi Z\alpha\,\delta^{d}(r_1)\bigg(E_0-V-\frac{p_2^2}{2}\bigg)\nonumber\\
&&+\frac{2}{3}
\bigg[\frac{(Z\alpha)^2}{r_1^4}\bigg]_\epsilon-\frac{4}{3}
\bigg[\frac{Z\alpha\,\vec r_1}{r_1^3}\bigg]_\epsilon\cdot\bigg[\frac{\alpha\,\vec r}{r^3}\bigg]_\epsilon
+\frac{1}{20}p_1^i\,4\pi Z\alpha\,\delta^{d}(r_1)\,p_1^i
\bigg)
+(1\leftrightarrow2)\bigg\}\nonumber\\
&&+\alpha^2\bigg\{-\frac{68}{45}\big(\vec P_1-\vec P_2\big)^2-\frac{62}{225} q^2
-\frac{136}{45}\vec P_1\cdot\vec P_2
-\frac{112}{45}\frac{\big(\vec P_1\cdot\vec q\big)\big(\vec P_2\cdot\vec q\big)}{q^2}\nonumber\\
&&+\ln\Lambda_{\epsilon}\bigg(
-\frac{2}{3}\big(\vec P_1-\vec P_2\big)^2-\frac{2}{15} q^2 -\frac{4}{3}\vec P_1\cdot\vec P_2
-\frac{4}{3}\frac{\big(\vec P_1\cdot\vec q\big)\big(\vec P_2\cdot\vec q\big)}{q^2}\bigg)
\bigg\}
\nonumber\\
&&
+\pi\alpha^3\bigg(-\frac{103}{45}-\frac43\ln\Lambda_{\epsilon}-\frac83\ln2+\frac83\ln q\bigg)\,q
\,\bigg|\phi\bigg\rangle
\,.
\end{eqnarray}

\subsubsection{Total low-energy result}

Adding together $E^\Lambda_{L1}$, $E^\Lambda_{L2}$, and $E^\Lambda_{L3}$, we arrive at the final
result for the low-energy contribution, which is
\begin{align}\label{LEtot2}
H_{L}^\Lambda
=&\
\frac{\alpha}{\pi}
\bigg\{\pi Z\alpha\,\delta^{d}(r_1)\bigg[
-\frac{161}{225}\bigg(E_0-V-\frac{p_2^2}{2}\bigg)
-\frac{11}{15}\ln\Lambda_\epsilon \bigg(E_0-V-\frac{p_2^2}{2}\bigg)\bigg]
+\frac{53}{90}\bigg[\frac{(Z\alpha)^2}{r_1^4}\bigg]_\epsilon-\frac{53}{45}
\bigg[\frac{Z\alpha\,\vec r_1}{r_1^3}\bigg]_\epsilon\cdot\bigg[\frac{\alpha\,\vec r}{r^3}\bigg]_\epsilon\nonumber\\
&\
+\frac{341}{450}p_1^i\pi Z\alpha\,\delta^{d}(r_1)\,p_1^i
+\ln\Lambda_\epsilon\bigg(\frac{1}{3}\bigg[\frac{(Z\alpha)^2}{r_1^4}\bigg]_\epsilon
-\frac{2}{3}\bigg[\frac{Z\alpha\,\vec r_1}{r_1^3}\bigg]_\epsilon\cdot\bigg[\frac{\alpha\,\vec r}{r^3}\bigg]_\epsilon
+\frac{11}{30}p_1^i\,\pi Z\alpha\,\delta^{d}(r_1)\,p_1^i\bigg)
+(1\leftrightarrow2)\bigg\}\nonumber\\&\
+\pi\alpha^3\bigg\{-\frac{73}{45}-\frac43\ln\Lambda_{\epsilon}-\frac{8}{3}\ln2+\frac{8}{3}\ln q
\bigg\}\,q
+\alpha^2\bigg\{-\frac{2}{5}\big(\vec P_1-\vec P_2\big)^2
+\frac{188}{225} q^2
+\frac{64}{45}\vec P_1\cdot\vec P_2
\nonumber\\&\
-\frac{4}{15}\frac{\big(\vec P_1\cdot\vec q\big)\big(\vec P_2\cdot\vec q\big)}{q^2}
+\ln\Lambda_{\epsilon}\bigg(
\frac{4}{3}\vec P_1\cdot\vec P_2+\frac{8}{15} q^2\bigg)
+\sigma_1\cdot\sigma_2\bigg(\frac{7}{27}+\frac19\ln\Lambda_{\epsilon}\bigg)q^2\bigg\}\,.
\end{align}

\subsection{Middle-energy part}

We now turn to the second term in Eq.~(\ref{eq:2}), the middle-energy contribution $E_{M}$. This
part originated from the region where both the radiative and the exchanged photons are of the
order $k \propto m\alpha$. We separate $E_{M}$ into two parts, $E_{M} = E_{M1} + E_{M2}$, which are
examined as follows.

\subsubsection{Triple seagull contribution}

The first middle-energy part is the triple seagull contribution, which is expressed (with $k_3$
being the radiative photon) as
\begin{eqnarray}
E_{M1} &=& e^6\int\frac{d^d k_1}{(2\pi)^d\,2k_1}
\int\frac{d^d k_2}{(2\pi)^d\,2k_2}\int\frac{d^d k_3}{(2\pi)^d\,2k_3}
\,\delta_\perp^{ik}(k_1)\delta_\perp^{jk}(k_2)\delta_\perp^{ij}(k_3)
 \nonumber\\&&  \times
\bigg<\phi\bigg| e^{{\rm i}(\vec k_1+\vec k_2)\cdot\vec r_1}\,\frac{1}{E_0-H_0-k_1-k_2}\,e^{{\rm i}(\vec k_3-\vec k_1)\cdot\vec r_2}\,\frac{1}{E_0-H_0-k_2-k_3}\,e^{-{\rm i}(\vec k_2+\vec k_3)\cdot\vec r_2}\nonumber\\
&&+e^{-{\rm i}(\vec k_1+\vec k_3)\cdot\vec r_2}\,\frac{1}{E_0-H_0-k_1-k_3}\,e^{-{\rm i}(\vec k_2-\vec k_3)\cdot\vec r_2}\,\frac{1}{E_0-H_0-k_1-k_2}\,e^{{\rm i}(\vec k_1+\vec k_2)\cdot\vec r_1}\nonumber\\
&&+e^{-{\rm i}(\vec k_1+\vec k_3)\cdot\vec r_2}\,\frac{1}{E_0-H_0-k_1-k_3}\,e^{{\rm i}(\vec k_1+\vec k_2)\cdot\vec r_1}\,\frac{1}{E_0-H_0-k_2-k_3}\,e^{-{\rm i}(\vec k_2+\vec k_3)\cdot\vec r_2}
 \bigg|\phi\bigg>
+(1\leftrightarrow2)\,.
\end{eqnarray}
We now have to expand the resolvents for large $k$. To get the contribution of the order
$\alpha^7m$ it is sufficient to take the nonretardation approximation, thus omitting $H_0-E_0$.
We arrive at
\begin{eqnarray}
E_{M1}&=& e^6\int\frac{d^d k_1}{(2\pi)^d\,2k_1}
\int\frac{d^d k_2}{(2\pi)^d\,2k_2}\int\frac{d^d k_3}{(2\pi)^d\,2k_3}
\,\delta_\perp^{ik}(k_1)\delta_\perp^{ik}(k_2)\frac{(d-1)}{d}
 \nonumber\\&& \times
\bigg<\phi\bigg|e^{{\rm i}(\vec k_1+\vec k_2)\cdot\vec r}\bigg[\frac{1}{(k_1+k_2)(k_2+k_3)}+\frac{1}{(k_1+k_3)(k_1+k_2)}+\frac{1}{(k_1+k_3)(k_2+k_3)}\bigg]
\bigg|\phi\bigg>
+(1\leftrightarrow2)\,.
\end{eqnarray}
Similarly to the case of the low-energy contribution, we will express $E_{M1}$ as an expectation
value of some effective operator $H_{M1}$. Because it contains purely two-body electron-electron
terms, we express it in momentum representation. We obtain
\begin{eqnarray}
H_{M1}&=& (4\pi\alpha)^3\frac{(d-1)}{8d}
\int\frac{d^d k}{(2\pi)^d}\int\frac{d^d k_3}{(2\pi)^d}
\,\frac{\delta_\perp^{ik}(k_1)\delta_\perp^{ik}(k_2)}{k_1\, k_2\, k_3}
 \nonumber\\&& \times
\bigg[\frac{1}{\big(k_1+k_2\big)\big(k_2+k_3\big)}
+\frac{1}{\big(k_1+k_3\big)\big(k_1+k_2\big)}
+\frac{1}{\big(k_1+k_3\big)\big(k_2+k_3\big)}\bigg]+(1\leftrightarrow2)\,,
\end{eqnarray}
with $k_1 = \big|\vec k-\frac{\vec q}{2}\big|$ and $k_2=\big|\vec k+\frac{\vec q}{2}\big|$. The integration over
radiative photon $k_3$ is trivial. The remaining integration is performed in spheroidal
coordinates as explained in Appendix \ref{app:spheroid}. The result for the triple seagull
contribution is
\begin{equation}
H_{M1} = \alpha^3\pi\bigg(-\frac13+\frac43\ln2\bigg)\,q.
\end{equation}

\subsubsection{Single seagull with retardation}

The second middle-energy contribution comes from the diagram with a single seagull and
retardation. The diagram contains two photons, one of which is a transverse photon exchanged
between the electrons and another one is a radiative photon.
The corresponding contribution is expressed as
\begin{eqnarray}
E_{M2} &=&e^4\int\frac{d^d k_1}{(2\pi)^d\,2k_1}\int\frac{d^d k_2}{(2\pi)^d\,2k_2}\delta_\perp^{in}(k_1)\delta_\perp^{im}(k_2)
 \nonumber\\ & \times&
\bigg<\phi\bigg|
j_1^n(k_1)\,e^{{\rm i}\vec k_1\cdot\vec r_1}\,\frac{1}{E_0-H_0-k_1}\,e^{-{\rm i}(\vec k_1+\vec k_2)\cdot\vec r_2}\,\frac{1}{E_0-H_0-k_2}\,j_2^m(k_2)\,e^{{\rm i}\vec k_2\cdot\vec r_2}\nonumber\\
&&+j_2^n(k_1)\,e^{{\rm i}\vec k_1\cdot\vec r_2}\,\frac{1}{E_0-H_0-k_1}\,e^{-{\rm i}(\vec k_1+\vec k_2)\cdot\vec r_2}\,\frac{1}{E_0-H_0-k_2}\,j_1^m(k_2)\,e^{{\rm i}\vec k_2\cdot\vec r_1}\nonumber\\
&&+j_1^n(k_1)\,e^{{\rm i}\vec k_1\cdot\vec r_1}\,\frac{1}{E_0-H_0-k_1}\,j_2^m(k_2)\,e^{{\rm i}\vec k_2\cdot\vec r_2}\,\frac{1}{E_0-H_0-k_1-k_2}\,e^{-{\rm i}(\vec k_1+\vec k_2)\cdot\vec r_2}\nonumber\\
&&+j_2^n(k_1)\,e^{{\rm i}\vec k_1\cdot\vec r_2}\,\frac{1}{E_0-H_0-k_1}\,j_1^m(k_2)\,e^{{\rm i}\vec k_2\cdot\vec r_1}\,\frac{1}{E_0-H_0-k_1-k_2}\,e^{-{\rm i}(\vec k_1+\vec k_2)\cdot\vec r_2}\nonumber\\
&&+e^{-{\rm i}(\vec k_1+\vec k_2)\cdot\vec r_2}\,\frac{1}{E_0-H_0-k_1-k_2}\,j_1^n(k_1)\,e^{{\rm i}\vec k_1\cdot\vec r_1}\,\frac{1}{E_0-H_0-k_2}\,j_2^m(k_2)\,e^{{\rm i}\vec k_2\cdot\vec r_2}\nonumber\\
&&+e^{-{\rm i}(\vec k_1+\vec k_2)\cdot\vec r_2}\,\frac{1}{E_0-H_0-k_1-k_2}\,j_2^n(k_1)\,e^{{\rm i}\vec k_1\cdot\vec r_2}\,\frac{1}{E_0-H_0-k_2}\,j_1^m(k_2)\,e^{{\rm i}\vec k_2\cdot\vec r_1}
\bigg|\phi\bigg>
+(1\leftrightarrow2)\,,
\end{eqnarray}
where $j_i^l(k)$ is defined as
\begin{equation}
j_i^l(k) = p^l_i + \frac{{\rm i}}{2}\sigma_i^{kl} k^k.
\end{equation}
The first two terms in the matrix element vanish after performing the retardation expansion and
carrying out the integration over the momentum of the radiative photon as $ \int d^d k\,k^\alpha
= 0$, which is true by definition in the dimensional regularization. The remainder of the
expression can be cast in the form
\begin{eqnarray}
E_{M2} &=&-2e^4\int\frac{d^d k_1}{(2\pi)^d\,2k_1}\int\frac{d^d k_2}{(2\pi)^d\,2k_2}\frac{1}{k_1^2(k_1+k_2)}
\,\delta_\perp^{in}(k_1)\delta_\perp^{im}(k_2)\nonumber\\
&&
\times
\bigg<\phi\bigg|
\bigg[\bigg[j_1^n(k_1)\,e^{{\rm i}\vec k_1\cdot\vec r_1},H_0-E_0\bigg],j_2(k_2)\,e^{{\rm i}\vec
k_2\cdot\vec r_2}\bigg]\,e^{-{\rm i}(\vec k_1+\vec k_2)\cdot\vec r_2}
\bigg|\phi\bigg>
+(1\leftrightarrow2)\,.
\end{eqnarray}
\end{widetext}
The above expression was rearranged so that only photon $k_2$ is radiative (thus the additional
factor of 2 in the front). Taking into account that only spin-independent terms survive the
double commutator and performing the angular average for the radiative photon, we arrive at
\begin{align}
E_{M2} =&\, -\frac{(4\pi\alpha)^2}{2}\frac{(d-1)}{d}\int\frac{d^d k_1}{(2\pi)^d}
\int\frac{d^d k_2}{(2\pi)^d}\frac{1}{k_1^3k_2(k_1+k_2)}
 \nonumber \\ & \times
\delta_\perp^{mn}(k_1) \langle\phi|\,e^{{\rm i} \vec k_1\cdot\vec r}\,\partial_1^m\partial_2^n
V|\phi\rangle +(1\leftrightarrow2)\,.
\end{align}
This can be again expressed as an expectation value of an effective operator $H_{M2}$, which is
in momentum space
\begin{align}
H_{M2} =&\, -(4\pi\alpha)^3\frac{(d-1)}{d}\int\frac{d^d k_1}{(2\pi)^d}
\int\frac{d^d k_2}{(2\pi)^d}
 \nonumber \\ & \times
\frac{\delta_\perp^{mn}(k_1) q^m q^n}{k_1^3\,|\vec{k}_1-\vec{q}|^2\,k_2(k_1+k_2)} \,.
\end{align}
Performing the remaining integrations, we obtain the result
\begin{equation}
H_{M2} = \alpha^3\pi\bigg(\frac49+\frac{2}{3\epsilon}-\frac83\ln q\bigg)\,q\,.
\end{equation}

\subsubsection{Total result for the middle-energy contribution}

Adding together $H_{M1}$ and $H_{M2}$, we obtain the total result for the effective operator
responsible for the middle-energy contribution,
\begin{equation} \label{eq:EM}
H_M = \pi\alpha^3\,\bigg(\frac19+\frac{2}{3\epsilon}+\frac43\ln2-\frac83\ln q\bigg)\,q\,.
\end{equation}
It is not obvious that $E_M=\langle H_M\rangle$ is the complete middle-energy contribution, so we
have verified this by calculating the corresponding scattering amplitude and obtaining agreement
with the above result.

\subsection{High-energy part}
\label{sec:EH}

We now turn to the third term in Eq.~(\ref{eq:2}), the high-energy part $E_{H}$. It consists of
16 terms originating from the anomalous magnetic moment $\kappa$ and the slopes of form factors
$F_1'(0)$ and $F_2'(0)$,
\begin{equation}
 E_H = \sum_{i = 1}^{16} E_{i}\,.
\end{equation}
The contributions $E_{i}$ are derived as corrections to the one-photon and two-photon exchange
amplitudes induced by individual terms in the NRQED Hamiltonian $H_{\rm nrqed}$, see
Eq.~(\ref{eq:hrel}), as illustrated by Table~\ref{tab:Ei1}. The computational method is very
similar to the one used in the derivation of the $\alpha^6\,m$ correction to the helium Lamb
shift \cite{he_singlet}. Each contribution $E_i$ will be expressed as an expectation value of the
corresponding effective operator, $E_i = \langle H_i \rangle$, and we now examine the
contributions $E_i$ one by one.

\begin{table*}
\caption{
  Contributions originating from $\delta H_1$, $\delta H_2$ and $\delta H_3$, given by Eqs.~(\ref{trafo3}),
	(\ref{trafo3b}) and (\ref{deltaH}).
	\label{tab:Ei1}}
\begin{ruledtabular}
  \begin{tabular}{LLLLL}
\centt{vertex} &\centt{vertex} & \centt{vertex} &\centt{retardation order} &\centt{diagram}\\[2pt] \hline
\hline
  -\frac{e}{4}\sigma\cdot B&
  -\frac{e}{4}\kappa\,\sigma\cdot B&&
  (H-E)^2&E_1\\[8pt]
  \frac{e^2}{2}\vec{A}\,{}^2&
  -\frac{e}{4}\kappa\,\sigma\cdot B&
  -\frac{e}{4}\sigma\cdot B&(H-E)^0&E_2\\[8pt]
   \frac{e}{16}\bigl\{\vec{p}\,{}^2,\,\sigma\cdot B\bigr\} &
-\frac{e}{4}\kappa\,\sigma\cdot B &&(H-E)^0&E_3\\[8pt]
		-\frac{e}{4}\sigma\cdot B&
   \frac{e}{16}\kappa\,\big(\frac12\{\vec{p}\,{}^2,\sigma\cdot B\}+4p^k\sigma^{ki} B^{ij}\,p^j
	 + \vec p\,\sigma\cdot B\,\vec p\big)&&(H-E)^0&E_4\\[8pt]
  \frac{e^2}{4}\sigma^{ij}{E}_\parallel^i A^j&
  -\frac{e}{4}\kappa\,\sigma\cdot B & e\,A^0 &(H-E)^0&E_5\\[8pt]
	  -\frac{e}{4}\sigma\cdot B&
  \frac{e^2}{2}\kappa\,\sigma^{ij}{E}_\parallel^i A^j & e\,A^0 &(H-E)^0&E_6\\[8pt]
	  -\frac{e}{4}\sigma\cdot B&
  -\frac{e}{4m^2}\kappa\,\sigma^{ij}\{E_\perp^i,p^j\}&&
  (H-E)&E_7\\[8pt]
  -\frac{e}{8}\,\sigma^{ij}\{E_\perp^i,p^j\} &
	  -\frac{e}{4}\kappa\,\sigma\cdot B &&
  (H-E)&E_{8}\\[8pt]
	e A^0 & -\frac{e}{8}\big(F_1'(0)+2F_2'(0)+4F_1''(0)\big)\vec\nabla^2\vec\nabla\cdot\vec E & & (H-E)^0 &E_{10}\\[8pt]
e A^0 & \frac{1}{16}\kappa\,\{\vec p^2,\vec\nabla\cdot\vec E\} & &\,(H-E)^0 &E_{11}\\[8pt]
 e A^0 &  \frac{e^2}{8}\kappa\, \vec E_\parallel^2 & e A^0 &\,(H-E)^0& E_{12}\\[8pt]
 e A^0 &  e^2\,\vec E_\parallel^{2}\,\chi &e A^0  &\,(H-E)^0 & E_{13}\\[8pt]
	-\frac{e}{8}\,
\left(\vec\nabla\cdot\vec E+\sigma^{ij}\,\{E_\parallel^i,p^j\}\right)&
-\frac{e}{4}\,\kappa\,
\left(\vec\nabla\cdot\vec E+\sigma^{ij}\,\{E_\parallel^i,p^j\}\right)&
&(H-E)^0& E_{14}\\[8pt]
-e \,\vec{p}\cdot\vec{A}&
 -\frac{e\,\kappa}{16} \,
         \{p^k,\partial_t E^k_\perp + \nabla^i\,B^{ik}\} &&
(H-E)^0 & E_{15}\\[8pt]
e A^0&- \frac{e\,\kappa}{16} \,
         \{p^k, \partial_{t}E_\parallel^k \} &
				 &(H-E)^0
 & E_{16}\\[8pt]
\end{tabular}
\end{ruledtabular}
\end{table*}

\subsubsection{$E_1$}

$E_1$ is the retardation correction to the one-photon exchange between the electrons, where one
vertex is $-\frac{e}{4}\sigma\cdot B$ and the second
vertex is $ -\frac{e}{4}\kappa\,\sigma\cdot B$. We have
\begin{align}
E_1 =&\  -e^2\kappa \int \frac{d^dk}{(2\pi)^d\,2k^4}\delta_\perp^{ij}(k)
\langle\phi|\biggl(\frac{1}{2}\sigma_1^{ki}\nabla_1^k\biggr)e^{{\rm i}\vec{k}\cdot\vec{r}_1}
 \nonumber \\ & \times
(H_0-E_0)^2
\biggl(\frac{1}{2}\sigma_2^{lj}\nabla_2^l\biggr)e^{-{\rm i}\vec{k}\cdot\vec{r}_2}|\phi\rangle
+\textrm{h.c.} + (1\leftrightarrow2)\,.
\end{align}
Commuting $H_0-E_0$ to the left and to the right and performing the spin averaging, we obtain
\begin{align}
E_1 =&\  -\frac{e^2}{4}\kappa\, \frac{\sigma_1\cdot\sigma_2}{d}\int \frac{d^dk}{(2\pi)^d k^2}
 \nonumber \\ & \times
 \langle\phi|\bigg[\frac{p_2^2}{2}\bigg[e^{{\rm i}\vec{k}\cdot\vec{r}},\frac{p_1^2}{2}\bigg]\bigg]|\phi\rangle + (1\leftrightarrow2)\,.
\end{align}
We now use the result for $\kappa \equiv F_2(0)$ from Appendix~\ref{app:ffactors},
\begin{equation}
\kappa = \frac{\alpha}{\pi}\bigg[\frac12+2\epsilon\bigg]\,,
\end{equation}
and transform the momenta into $\vec P_1$ and $\vec P_2$.
We obtain in the momentum representation
\begin{equation}
H_1 = -\frac{\alpha^2}{3}\sigma_1\cdot\sigma_2\,\frac{(\vec P_1\cdot \vec q)(\vec P_2\cdot \vec q)}{q^2}\,.
\end{equation}

\subsubsection{$E_2$}

$E_2$ is induced by the single-seagull diagram with the double vertex
$\frac{e^2}{2}\vec{A}\,{}^2$ and the single vertices $ -\frac{e}{4}\sigma\cdot B$ and $-\frac{e}{4}\kappa\,\sigma\cdot B$. The corresponding
contribution is written as
\begin{widetext}
\begin{eqnarray}
E_2 &=&
e^4\kappa\, \int \frac{d^dk_1}{(2\pi)^d\,2k_1} \int \frac{d^dk_2}{(2\pi)^d\,2k_2}
\delta_\perp^{in}(k_1)\delta_\perp^{im}(k_2)\nonumber\\
&&\times
\biggl\{\langle\phi|\bigg(\frac{1}{2}\sigma_1^{rn}\nabla_1^r\bigg)
\,e^{{\rm i}\vec{k}_1\cdot\vec{r}_1}\,\frac{1}{E_0-H_0-k_1}\,e^{-{\rm i}(\vec{k}_1+\vec{k}_2)\cdot\vec{r}_2}\,
\frac{1}{E_0-H_0-k_2}\,\bigg(\frac{1}{2}\sigma_1^{sm}\nabla_1^s\bigg)\,e^{{\rm i}\vec{k}_2\cdot\vec{r}_1}|\phi\rangle\nonumber\\
&&+\,
\langle\phi|\bigg(\frac{1}{2}\sigma_1^{rn}\nabla_1^r\bigg)\,e^{{\rm i}\vec{k}_1\cdot\vec{r}_1}\,\frac{1}{E_0-H_0-k_1}
\,\bigg(\frac{1}{2}\sigma_1^{sm}\nabla_1^s\bigg)\,e^{{\rm i}\vec{k}_2\cdot\vec{r}_1}\,
\frac{1}{E_0-H_0-k_1-k_2}\,e^{-{\rm i}(\vec{k}_1+\vec{k}_2)\cdot\vec{r}_2}|\phi\rangle\nonumber\\
&&+\,
\langle\phi|e^{-{\rm i}(\vec{k}_1+\vec{k}_2)\cdot\vec{r}_2}\,\frac{1}{E_0-H_0-k_1-k_2}
\,\bigg(\frac{1}{2}\sigma_1^{rn}\nabla_1^r\bigg)\,e^{{\rm i}\vec{k}_1\cdot\vec{r}_1}\,
\frac{1}{E_0-H_0-k_2}\,\bigg(\frac{1}{2}\sigma_1^{sm}\nabla_1^s\bigg)\,e^{{\rm i}\vec{k}_2\cdot\vec{r}_1}|\phi\rangle\biggr\} + (1\leftrightarrow2)\,.
\end{eqnarray}
Expanding denominators in large $k$, we get the $\alpha^7m$ contribution of the form
\begin{eqnarray}
E_2 &=&
\frac{e^4}{2}\kappa\int \frac{d^dk_1}{(2\pi)^d\,k_1^2} \int \frac{d^dk_2}{(2\pi)^d\,k_2^2}
\delta_\perp^{in}(k_1)\delta_\perp^{im}(k_2)
\langle\phi|\bigg(\frac{{\rm i}}{2}\sigma_1^{rn}k_1^r\bigg)\,e^{{\rm i}(\vec{k}_1+\vec k_2)\cdot\vec{r}}
\,\bigg(\frac{{\rm i}}{2}\sigma_1^{sm}k_2^s\bigg)|\phi\rangle
 + (1\leftrightarrow2)\nonumber\\
&=&
-\frac{e^4}{8}\kappa\int \frac{d^dk_1}{(2\pi)^d\,k_1^2} \int \frac{d^dk_2}{(2\pi)^d\,k_2^2}\sigma_1^{rn}\sigma_1^{sm}k_1^r k_2^s\,
\delta_\perp^{in}(k_1)\delta_\perp^{im}(k_2)
\langle\phi|\,e^{{\rm i}(\vec{k}_1+\vec k_2)\cdot\vec{r}}|\phi\rangle
 + (1\leftrightarrow2)
 \,.
\end{eqnarray}
Averaging the spin matrices and performing the integrations by using formulas from Appendix~C of
Ref.~\cite{patkos:20}, we obtain the result for $E_2$ in the momentum representation,
\begin{equation}
H_2 = -\pi\alpha^3\,\frac{q}{4}\,.
\end{equation}

\subsubsection{$E_3$}

$E_3$ is induced by the exchange of a single transverse photon, with one vertex
$-\frac{e}{4}\kappa\,\sigma\cdot B$ and the other
   $\frac{e}{4}\bigl\{\vec{p}\,{}^2,\frac14\,\sigma\cdot B\bigr\}$.
We thus get
\begin{eqnarray}
E_3 &=& \frac{e^2}{4}\kappa\int\frac{d^dk}{(2\pi)^d \,2k^2}\delta_\perp^{ij}(k)
\langle\phi|\biggl(\frac{1}{2}\sigma_1^{ki}\nabla_1^k\biggr)\,e^{{\rm i}\vec{k}\cdot\vec{r}_1}
\,\biggl\{p_2^2,\frac12\sigma_2^{lj}\nabla_2^l\,e^{-{\rm i}\vec{k}\cdot\vec{r}_2}\biggr\}\,|\phi\rangle
+\textrm{h.c.} + (1\leftrightarrow2)\nonumber\\
&=& \frac{e^2}{32}\kappa\frac{\sigma_1\cdot\sigma_2}{d}\int\frac{d^dk}{(2\pi)^d}
\langle\phi|\bigl\{p_2^2,e^{{\rm i}\vec{k}\cdot\vec{r}}\bigr\}\,|\phi\rangle
+\textrm{h.c.} + (1\leftrightarrow2)
\,.
\end{eqnarray}
This expression is proportional to $\{p_2^2,\delta^d(r)\}$, which vanishes for
triplet states. Therefore, $E_3 = 0\,$.

\subsubsection{$E_4$}

$E_4$ comes from the exchange of a single transverse photon, with one vertex being
$-\frac{e}{4}\sigma\cdot B$ and the other  $
\frac{e}{16}\kappa\,\big(\frac12\{\vec{p},\{\vec p,\sigma\cdot B\}\}+4p^k\sigma^{ki}
B^{ij}\,p^j\big)$. The corresponding contribution is written as
\begin{eqnarray}
E_4 &=& \frac{e^2}{16}\kappa\int\frac{d^dk}{(2\pi)^d \,2k^2}\delta_\perp^{ij}(k)
\langle\phi|\biggl(\frac{1}{2}\sigma_1^{ki}\nabla_1^k\biggr)\,e^{{\rm i}\vec{k}\cdot\vec{r}_1}
 \nonumber\\&&
 \times
\biggl[\{p_2^l,\{p_2^l,\sigma_2^{mj}\nabla_2^m\,e^{-{\rm i}\vec{k}\cdot\vec{r}_2}\}\}
+4p_2^l\bigg(\sigma_2^{lm}\nabla_2^m \,e^{-{\rm i}\vec{k}\cdot\vec{r}_2}\,p_2^j -\sigma_2^{lj}
\nabla_2^m \,e^{-{\rm i}\vec{k}\cdot\vec{r}_2}\,p_2^m\bigg)\biggr]\,|\phi\rangle
+\textrm{h.c.} + (1\leftrightarrow2)
\,.
\end{eqnarray}
After differentiating and spin averaging, we arrive at
\begin{eqnarray}
E_4 &=&
\frac{e^2}{16}\kappa\frac{\sigma_1\cdot\sigma_2}{d}\int\frac{d^dk}{(2\pi)^d \,k^2}
\langle\phi|\frac14\{\vec p_2,\{\vec p_2,k^2\,e^{{\rm i}\vec{k}\cdot\vec{r}}\}\}
-\frac{\delta_\perp^{ij}(k)}{(d-1)}\,p_2^i\,k^2\,e^{{\rm i}\vec{k}\cdot\vec{r}}\,p_2^j
-p_2^k\,k^k k^m  \,e^{{\rm i}\vec{k}\cdot\vec{r}}\,p_2^m\,|\phi\rangle
+\textrm{h.c.} + (1\leftrightarrow2)\,.\nonumber\\
\end{eqnarray}
Transforming the momenta in this expression into $\vec P_1$ and $\vec P_2$ we obtain the result in the momentum space,
\begin{equation}
H_4 = \alpha^2\sigma_1\cdot\sigma_2\bigg(\frac{1}{24}(\vec P_1-\vec P_2)^2+\frac{1}{12}\vec P_1\cdot\vec P_2
-\frac{1}{24}\frac{[(\vec P_1-\vec P_2)\cdot\vec q]^2}{q^2}+\frac{1}{24}q^2
-\frac{1}{12}\frac{(\vec P_1\cdot \vec q)(\vec P_2\cdot \vec q)}{q^2}\bigg)\,.
\end{equation}

\subsubsection{$E_5$ and $E_6$}

The terms $E_5$ and $E_6$ are examined together, because they both are given by the one-photon exchange amplitude
with one vertex $-\frac{1}{4}\sigma\cdot B$ and the other vertex
 $ e^2\,\sigma^{ij}{E}_\parallel^i A^j$, multiplied by a factor $\kappa/2$ in the case of
$E_6$ and $\kappa/4$ in the case of $E_5$. The sum of the contributions is
\begin{eqnarray}
  E_5+E_6 &=&-\frac{3e^2}{8}\kappa\int\frac{d^dk}{(2\pi)^d \,2k^2}\delta_\perp^{ij}(k)
\langle\phi|\sigma_1^{ki}\nabla_1^k\,e^{{\rm i}\vec{k}\cdot\vec{r}}
\,\sigma_2^{lj}\,\nabla_2^l\,V|\phi\rangle
+\textrm{h.c.} + (1\leftrightarrow2)\nonumber \\
&=&-\frac{3e^2}{16}\kappa\,\frac{\sigma_1\cdot\sigma_2}{d}
\int\frac{d^dk}{(2\pi)^d}\,\frac{k^l}{k^2}
\langle\phi|\,e^{{\rm i}\vec{k}\cdot\vec{r}} \,[V,p_2^l]|\phi\rangle
+\textrm{h.c.} + (1\leftrightarrow2)
\,.
\end{eqnarray}
The result in a mixed coordinate and momentum representations is
\begin{equation}
H_5+H_6 = \sigma_1\cdot\sigma_2
\bigg[\bigg\{-\frac{\alpha}{16\pi}\bigg[\frac{Z\alpha\,\vec r_1}{r_1^3}\bigg]_\epsilon
\cdot\bigg[\frac{\alpha\,\vec r}{r^3}\bigg]_\epsilon
+(1\leftrightarrow2)\bigg\}-\frac{\pi\alpha^3}{8}q\bigg]\,.
\end{equation}

\subsubsection{$E_7$ and $E_{8}$}

The terms $E_7$ and $E_{8}$ are induced by the one-photon exchange amplitude with one vertex
  $-\frac{e}{4}\sigma\cdot B$ and the other vertex
  $-e\kappa\,\sigma^{ij}\{E_\perp^i,p^j\}$, multiplied by either $\kappa/4$ or $\kappa/8$.
	The sum of the corresponding contributions is
\begin{eqnarray}
E_7+E_{8} &=& \frac{3e^2}{8}\kappa\int\frac{d^dk}{(2\pi)^d \,2k^3}\delta_\perp^{ij}(k)
\langle\phi|\sigma_1^{ik}\{{\rm i}k\,e^{{\rm i}\vec{k}\cdot\vec{r}_1},p_1^k\}\,(H_0-E_0)\,
\biggl(\frac{1}{2}\sigma_2^{lj}\nabla_2^l\biggr)\,e^{-{\rm i}\vec{k}\cdot\vec{r}_2}\,|\phi\rangle +\textrm{h.c.}+ (1\leftrightarrow2)\nonumber\\
&=& \frac{3e^2}{32}\kappa\,\sigma_1^{ik}\sigma_2^{lj}\int\frac{d^dk}{(2\pi)^d \,k^2}\delta_\perp^{ij}(k)\,
\langle\phi|\{e^{{\rm i}\vec{k}\cdot\vec{r}_1},p_1^k\}\,\bigg[H_0-E_0,\,
\bigg[e^{-{\rm i}\vec{k}\cdot\vec{r}_2},p_2^l\bigg]\bigg]\,|\phi\rangle
+\textrm{h.c.}+ (1\leftrightarrow2)\,.
\end{eqnarray}
Spin averaging this expression we get
\begin{eqnarray}
E_7+E_{8}
&=& -\frac{3e^2}{32}\kappa\,\frac{\sigma_1\cdot\sigma_2}{d}\int\frac{d^dk}{(2\pi)^d \,k^2}
\langle\phi|\,\bigg[\frac{p_2^2}{2},\,
\bigg[\{e^{{\rm i}\vec{k}\cdot\vec{r}},p_1^i\},p_2^i\bigg]\bigg]\,|\phi\rangle
+\textrm{h.c.}+ (1\leftrightarrow2)\,.
\end{eqnarray}
The result in momentum representation is
\begin{equation}
H_7+H_{8}  = \alpha^2\frac{\sigma_1\cdot\sigma_2}{2}\frac{(\vec P_1\cdot\vec q)(\vec P_2\cdot\vec q)}{q^2}
\,.
\end{equation}

\end{widetext}

\subsubsection{$E_{9}$}

We now move to contributions originating from the $\delta H_2$ part of the FW Hamiltonian, given
by Eq.~(\ref{trafo3b}). It is possible to evaluate them in the same way as the previous $E_i$
contributions, with the difference that one has to use the Feynman gauge instead of the Coulomb
gauge. This difference is caused by the presence of $q^2_0$ in the expansion of form factors $F_1$ and $F_2$. We
will use a different approach, however, which is more advantageous and illustrative, namely, the
so-called scattering amplitude approach. We recall that the electromagnetic form factors $F_1$
and $F_2$ modify the vertex $\gamma^\mu$ as
\begin{eqnarray}
\gamma^\mu \to \Gamma^\mu &=&
\gamma^\mu + \gamma^\mu\,F_1\big(q_0^2-q^2\big)\nonumber\\&& + \frac{{\rm i}}{2} F_2\big(q_0^2-q^2\big)
\left( {{\rm i}\over 2} \right)[\gamma^\mu, \qsla]\,.
\end{eqnarray}
The amplitude of the exchange of one photon between the electrons with one of the vertices
perturbed in this way is
\begin{equation}
M_{fi} = e^2 (\bar{u}'_1\Gamma^\mu u_1)D_{\mu\nu}(q)(\bar{u}'_2\gamma^\nu u_2)
+ (1\leftrightarrow2)\,,
\end{equation}
where $ q = p_1'-p_1 = p_2-p_2'$ with $p_1$ and $p_1'$ being the {\em in} and {\em out} momenta
of the first electron, and the same for the second electron. To obtain contributions up to the
order $\alpha^7m$, we expand the expression of the vertex as
\begin{eqnarray}
\Gamma^\mu &\to& \bigg\{\gamma^\mu\,\bigg(F'_1(0)-\frac12F''_1(0)\,q^2\bigg) - \frac{1}{4} F'_2(0)\,
q^j[\gamma^j, \gamma^\mu]\bigg\}\nonumber\\
&&\times\big[q_0^2-q^2\big]\,,
\end{eqnarray}
where we omitted terms with $q_0$ which contribute to higher orders in $\alpha$, and we also
excluded anomalous magnetic moment contribution $\propto F_2(0)$. We now pull the factor $q_0^2-q^2$
out of this expression to cancel it with the denominator of the photon propagator. The photon
propagator thus becomes
\begin{equation}
(q_0^2-q^2) D_{\mu\nu}(q) = g_{\mu\nu}
\,.
\end{equation}
The scattering amplitude is then
\begin{align}
M_{fi} =&\ e^2\bigg\{\Big(F_1'(0)-\frac12F_1''(0)\,q^2\Big)
 \nonumber \\ &\times
\big[(\bar{u}'_1\gamma^0 u_1)(\bar{u}'_2\gamma^0 u_2)-
(\bar{u}'_1\gamma^i u_1)(\bar{u}'_2\gamma^i u_2)\big]
 \nonumber\\&
 -\frac14 F_2'(0)\big[(\bar{u}'_1 q^j[\gamma^j, \gamma^0] u_1)(\bar{u}'_2\gamma^0 u_2)
 \nonumber\\&
 -
(\bar{u}'_1 q^j[\gamma^j, \gamma^i] u_1)(\bar{u}'_2\gamma^i u_2)\big]
+(1\leftrightarrow2)\bigg\}
\,.
\end{align}
For the bispinor $u$ we take
\begin{equation}
u =
\begin{pmatrix} \big(1-\frac{p^2}{8}\big)\,w \\ \frac{(\vec\sigma\cdot\vec p)}{2}\, w\end{pmatrix}\,,
\end{equation}
where $w$ is the spinor amplitude of the plane wave that includes the relativistic correction to
the kinetic energy. Then,
\begin{align}
&(\bar{u}'_1\gamma^0 u_1) =   (w_1')^*\Big(1-\frac{q^2}{8}+\frac{{\rm i}}{4}\sigma_1^{ij}q^i p^j_1\Big) w_1\,,
 \nonumber \\
&(\bar{u}'_1\gamma^i u_1) =  \frac12(w_1')^*\big({\rm i}\,\sigma_1^{ji}q^j+2p^i_1+q^i\big)\,w_1\,,
 \nonumber\\
&(\bar{u}'_1 q^j[\gamma^j, \gamma^0] u_1) = (w_1')^*\big( q^2 + 2\,{\rm i}\, \sigma_1^{ij} q^j p_1^i\big) w_1\,,
 \nonumber \\
&(\bar{u}'_1 q^j[\gamma^j, \gamma^i] u_1) = (w_1')^*\,2\,{\rm i}\, \sigma_1^{ij} q^j\,w_1 \,.
\end{align}
The corresponding expressions for the second electron are obtained from the above formulas by
changing the index $1\to 2$ and reversing the transferred momentum $\vec q\rightarrow-\vec q$. We
thus obtain for the scattering amplitude
\begin{eqnarray}
M_{fi} &=& -(w_1')^*(w_2')^* \,U(\vec p_1,\vec p_2,\vec q)\,w_1 w_2\,,
\end{eqnarray}
where (omitting higher-order terms and terms contributing only to the fine structure)
\begin{align}
U(\vec p_1,\vec p_2,\vec q) =&\ e^2 \bigg\{F_1'(0)\bigg[-1+\frac{q^2}{4}
+\frac14\bigg(\sigma_1^{ik}\sigma_2^{jk}\, q^i q^j
 \nonumber \\ &
+ (2\vec p_1+\vec q)(2\vec p_2-\vec q)\bigg)\bigg] + \frac12F_1''(0)\,q^2
\nonumber\\&
+ F_2'(0) \bigg[\frac{q^2}{4}+\frac14\sigma_1^{ik}\sigma_2^{jk}\, q^i q^j\bigg]+(1\leftrightarrow2)\bigg\}
\,.
\end{align}
Transforming the expression from momenta $\vec p_1$ and $\vec p_2$ to  $\vec P_1$ and $\vec P_2$ and
spin-averaging it using the identity
\begin{eqnarray}
\sigma_1^{ik}\sigma_2^{jk}\, q^i q^j &=&
\frac{\sigma_1\cdot\sigma_2}{d}\, q^2
\,,
\end{eqnarray}
we finally obtain
\begin{align}
U(\vec p_1,\vec p_2,\vec q) = &\ e^2 \bigg\{F_1'(0)\bigg[-1+\frac{q^2}{4} + \vec P_1\cdot\vec P_2
+\frac{\sigma_1\cdot\sigma_2}{4d}\, q^2\bigg]
 \nonumber\\ &
 + \frac12F_1''(0)\,q^2
+ F_2'(0) \bigg[\frac{q^2}{4}+\frac{\sigma_1\cdot\sigma_2}{4d}\, q^2\bigg]
 \nonumber\\ &
+(1\leftrightarrow2)\bigg\}
\,.
\end{align}
The first term in this equation corresponds to the leading order contribution while the remaining
ones are the $\alpha^7 m$ corrections. Using the explicit results for the form factors, we get
\begin{align}
H_{9} = &\ \alpha^2\bigg[-\frac{9}{20} q^2-\vec P_1\cdot\vec P_2
+\frac{1}{\epsilon}\bigg(-\frac{8}{15} q^2 - \frac43\vec P_1\cdot\vec P_2\bigg)
 \nonumber \\ &
+\sigma_1\cdot\sigma_2\bigg(-\frac{11}{108}-\frac{1}{9\epsilon}\bigg)q^2\bigg] \,.
\end{align}

\subsubsection{$E_{10}$}

Next, we have to take into account the exchange of a Coulomb photon between an electron and the
nucleus, originating from the Hamiltonian $\delta H_2$. This gives a correction in which one
vertex is $e A^0$ while the second one is
$-\frac{e}{8}\big[F_1'(0)+2F_2'(0)+4F_1''(0)\big]\vec\nabla^2\vec\nabla\cdot\vec E$. Only the
electron-nucleus part needs to be taken because the complete electron-electron contribution due to
derivative of the form factors was accounted for in $E_9$. So, the term $E_{10}$ is
\begin{align}
E_{10} &= \frac{1}{8}\Big[F_1'(0)+2F_2'(0)+4F_1''(0)\Big]\langle\phi|\vec\nabla_1^4 \bigg[-\frac{Z\alpha}{r_1}\bigg]_\epsilon|\phi\rangle
 \nonumber \\ &
+(1\leftrightarrow2)\,.
\end{align}
Using the relation (\ref{14}) and results for the form factors from Appendix~\ref{app:ffactors},
we evaluate it as (omitting linear in $\epsilon$ terms)
\begin{align}
H_{10} = &\ \alpha^2
\bigg(
-\frac{13}{40}-\frac{11}{30\epsilon}\bigg)\bigg\{-2\,\bigg(E_0-V-\frac{p_2^2}{2}\bigg)Z\,\delta^{d}(r_1)
 \nonumber \\ &
+\vec p_1\,Z\delta^{d}(r_1)\,\vec p_1 +(1\leftrightarrow2)\bigg\}\,.
\end{align}

\subsubsection{$E_{11}$}

We now continue to examine contributions originating from the exchange of the Coulomb photons.
The first such contribution, $E_{11}$, originates from the exchange of a Coulomb photon with one
vertex being $\frac{1}{16}\kappa\,\{\vec p{}^2,\vec\nabla\cdot\vec E\}$. The corresponding
contribution is
\begin{equation}
E_{11} = -\frac{1}{8}\kappa\,\langle\phi|\vec p_1{}^2\,\vec\nabla_1^2V|\phi\rangle
+(1\leftrightarrow2)\,.
\end{equation}
After some simplifications, we get
\begin{equation}
H_{11} = \alpha^2
\bigg[-\frac12\bigg(E_0-V-\frac{p_2^2}{2}\bigg)Z\,\delta^d(r_1)+(1\leftrightarrow2)\bigg]\,,
\end{equation}
where we omitted terms $p_1^2\,\delta^d(r)$ which vanish for triplet states.

\subsubsection{$E_{12}$ and $E_{13}$}
We treat the terms $E_{12}$ and $E_{13}$ together because they have similar structure.
Specifically, they both come from the exchange of a Coulomb photon where one vertex is either
$\frac{e^2}{8}\kappa\, \vec E_\parallel^2$ or $\delta H_3=e^2\,\vec
E_\parallel^{\;2}\,\chi$. We thus have
\begin{eqnarray}
E_{12}+E_{13} = \bigg(\frac\kappa8+\chi\bigg)\langle\phi|\big(\vec\nabla_1 V\big)^2|\phi\rangle
+(1\leftrightarrow2)\,.
\end{eqnarray}
Taking into account that
\begin{eqnarray}\label{nabla1V}
\big(\vec\nabla_1 V\big)^2 &=& \bigg[\frac{(Z\alpha)^2}{r_1^4}\bigg]_\epsilon
-2\,\bigg[\frac{Z\alpha\,\vec r_1}{r_1^3}\bigg]_\epsilon\cdot\bigg[\frac{\alpha\,\vec r}{r^3}\bigg]_\epsilon\nonumber\\
&&
-\pi^2 \alpha^2\big\{1+\epsilon\,(2\ln2-2\ln q)\big\}q\,,
\end{eqnarray}
  we obtain
\begin{align}
H_{12}+H_{13} = &\ \alpha\bigg(
\frac{11}{48}-\frac{1}{3\epsilon}+\frac14\epsilon\bigg)
\bigg\{\frac1\pi\bigg[\frac{(Z\alpha)^2}{r_1^4}\bigg]_\epsilon
 \nonumber \\ &
-\frac{2}{\pi}\bigg[\frac{Z\alpha\,\vec r_1}{r_1^3}\bigg]_\epsilon\cdot\bigg[\frac{\alpha\,\vec r}{r^3}\bigg]_\epsilon
+(1\leftrightarrow2)\bigg\}\nonumber\\ &
+\pi\alpha^3\,\bigg(-\frac{11}{24}+\frac{2}{3\epsilon}+\frac43\ln2-\frac43\ln q\bigg) q \,.
\end{align}

\subsubsection{$E_{14}$}

$E_{14}$ is induced by the exchange of a Coulomb photon with vertices
	$-\frac{e}{8}\, \big(\vec\nabla\cdot\vec E+\sigma^{ij}\,\{E_\parallel^i,p^j\}\big)$ and
$-\frac{e}{4}\,\kappa\, \big(\vec\nabla\cdot\vec E+\sigma^{ij}\,\{E_\parallel^i,p^j\}\big)$.
The corresponding contribution is
\begin{align}
E_{14} =& \ \frac{e^2}{32}\kappa\,\int\frac{d^dk}{(2\pi)^d\,k^2}\langle\phi|\big(k^2-2{\rm i}\sigma_1^{ij}k^i p_1^j\big)
 \nonumber \\ & \times
e^{{\rm i}\vec k\cdot\vec r}\,
\big(k^2+2{\rm i}\sigma_2^{kl}k^k p_2^l\big)|\phi\rangle
+(1\leftrightarrow2)\,.
\end{align}
Simplifying this expression, we get
\begin{equation}
H_{14} = \alpha^2\bigg[\frac18q^2+\sigma_1\cdot\sigma_2\bigg(\frac{1}{12}\vec P_1\cdot\vec P_2
-\frac{1}{12}\frac{(\vec P_1\cdot\vec q)(\vec P_2\cdot\vec q)}{q^2}\bigg)\bigg]\,.
\end{equation}

\begin{widetext}

\subsubsection{$E_{15}$}

$E_{15}$ originates from the one-photon  exchange with one vertex $-\frac{e\,\kappa}{16 } \,
         \{p^k, \partial_t E^k_\perp+\nabla^i\,B^{ik}\}$ and the other vertex $-e\,\vec{p}\cdot\vec A$.
We calculate this contribution starting from the corresponding Feynman diagram,
\begin{eqnarray}
E_{15} &=& \frac{e^2}{16} \kappa\int\frac{d^Dk}{(2\pi)^D {\rm i}}\frac{(-1)}{\omega^2-\vec{k}^2}\delta_\perp(k)^{ij}
\langle\phi|\{p^i_1,(\omega^2-\vec{k}^2) e^{{\rm i}\vec k\cdot\vec r_1}\}
\frac{1}{E_0-H_0-\omega+{\rm i}\epsilon} p_2^j\,e^{-{\rm i}\vec k\cdot\vec r_2}|\phi\rangle
+\textrm{h.c}+(1\leftrightarrow2)\nonumber\\
&=& -\frac{e^2}{16} \kappa\int\frac{d^Dk}{(2\pi)^D {\rm i}}\delta_\perp(k)^{ij}
\langle\phi|\{p^i_1,e^{{\rm i}\vec k\cdot\vec r_1}\}
\frac{1}{E_0-H_0-\omega+{\rm i}\epsilon} p_2^j\,e^{-{\rm i}\vec k\cdot\vec r_2}|\phi\rangle
+\textrm{h.c}+(1\leftrightarrow2)
\,.
\end{eqnarray}
Performing the $\omega$ integration and expressing the result in the momentum space, we obtain
\begin{eqnarray}
E_{15}
&=& \frac{e^2}{32}\kappa\int\frac{d^dk}{(2\pi)^d }\delta_\perp^{ij}(k)
\langle\phi|\big\{ p_1^i, e^{{\rm i}\vec k\cdot\vec r} \big\}
p_2^j\,|\phi\rangle +\textrm{h.c.}+ (1\leftrightarrow2)
 = \alpha^2\bigg\langle\frac12\vec P_1\cdot\vec P_2 -\frac12\frac{(\vec P_1\cdot \vec q)(\vec P_2\cdot \vec q)}{q^2}\bigg\rangle\,.
\end{eqnarray}

\subsubsection{$E_{16}$}

$E_{16}$ is induced by the exchange of a Coulomb photon with one vertex of the form $-
\frac{e\,\kappa}{16 } \,
         \{p^k, \partial_{t}E_\parallel^k \}$. The contribution from the corresponding Feynman diagram is
\begin{eqnarray}
E_{16} &=& \frac{e^2}{16} \kappa\int\frac{d^Dk}{(2\pi)^D {\rm i}}\bigg(\frac{-1}{\vec k^2}\bigg)
\bigg\{\langle\phi|\{\vec p_1,\vec k \omega e^{{\rm i}\vec k\cdot\vec r_1}\}
\frac{1}{E_0-H_0-\omega+{\rm i}\epsilon} e^{-{\rm i}\vec k\cdot\vec r_2}|\phi\rangle\nonumber\\
&&-\langle\phi|e^{-{\rm i}\vec k\cdot\vec r_2}\frac{1}{E_0-H_0-\omega+{\rm i}\epsilon}
\{\vec p_1,\vec k \omega e^{{\rm i}\vec k\cdot\vec r_1}\}|\phi\rangle\bigg\}
+(1\leftrightarrow2)\,,
\end{eqnarray}
where in the second term we introduced a mirror transformation $\vec k\rightarrow-\vec k$. The
denominator is now expanded for small $E_0-H_0$ up to the linear term and the $\omega$
integration is performed as
\begin{equation}
\int\frac{d\omega}{2\pi{\rm i}}\frac{\omega}{(\omega-{\rm i}\epsilon)^2} = \frac12.
\end{equation}
We then obtain
\begin{eqnarray}
E_{16} &=&-\frac{e^2}{32}\kappa \int\frac{d^dk}{(2\pi)^d k^2}\bigg\{\langle\phi|
\{\vec p_1,\vec k e^{{\rm i}\vec k\cdot\vec r_1}\}\bigg[\frac{p_2^2}{2},e^{-{\rm i}\vec k\cdot\vec r_2}\bigg]|\phi\rangle
-\langle\phi|\bigg[e^{-{\rm i}\vec k\cdot\vec r_2},\frac{p_2^2}{2}\bigg]
\{\vec p_1,\vec k  e^{{\rm i}\vec k\cdot\vec r_1}\}|\phi\rangle\bigg\}
+(1\leftrightarrow2)\nonumber\\
&=&-\frac{e^2}{32}\kappa \int\frac{d^dk}{(2\pi)^d k^2}\langle\phi|
\big[p_2^2,\{p^i_1, [p_1,e^{{\rm i}\vec k\cdot\vec r}]\}\big]|\phi\rangle
+(1\leftrightarrow2)
= \frac{\alpha^2}{2}\bigg\langle\frac{(\vec P_1\cdot \vec q)(\vec P_2\cdot \vec q)}{q^2}\bigg\rangle\,.
\end{eqnarray}

\subsubsection{Total high-energy part}

Adding together all $E_i$ contributions, we arrive at the final result for the high-energy
contribution,
\begin{eqnarray}\label{eq:EH}
E_H &=&
\bigg\langle \frac{\alpha}{\pi}\bigg\{\frac{11}{48}\bigg[\frac{(Z\alpha)^2}{r_1^4}\bigg]_\epsilon
-\bigg(\frac{11}{24}+\frac{\sigma_1\cdot\sigma_2}{16}\bigg)
\bigg[\frac{Z\alpha\,\vec r_1}{r_1^3}\bigg]_\epsilon\cdot\bigg[\frac{\alpha\,\vec r}{r^3}\bigg]_\epsilon
-\frac{13}{40}\vec p_1\,\pi Z\alpha\,\delta^d(r_1)\,\vec p_1
+ \frac{1}{\epsilon}\bigg(-\frac{1}{3}\bigg[\frac{(Z\alpha)^2}{r_1^4}\bigg]_\epsilon
\nonumber\\
&&+\frac23\bigg[\frac{Z\alpha\,\vec r_1}{r_1^3}\bigg]_\epsilon\cdot\bigg[\frac{\alpha\,\vec r}{r^3}\bigg]_\epsilon
-\frac{11}{30}\vec p_1\,\pi Z\alpha\,\delta^d(r_1)\,\vec p_1\bigg)
+\pi Z\alpha\,\delta^d(r_1)\bigg[\frac{3}{20} \bigg(E_0-V-\frac{p_2^2}{2}\bigg)
+\frac{11}{15\epsilon}\bigg(E_0-V-\frac{p_2^2}{2}\bigg)\bigg]\nonumber\\
&&+(1\leftrightarrow2)\bigg\}
 +\pi\alpha^3\bigg\{-\frac{17}{24}-\frac{\sigma_1\cdot\sigma_2}{8}+\frac{2}{3\epsilon} + \frac43\ln2-\frac43\ln q\bigg\}q
+\alpha^2\bigg\{- \frac12\vec P_1\cdot\vec P_2 - \frac{13}{40}q^2
+\frac1\epsilon\bigg(-\frac43\vec P_1\cdot\vec P_2-\frac{8}{15}q^2\bigg)
\nonumber\\
&&+\sigma_1\cdot\sigma_2\bigg[\frac{1}{24}(\vec P_1-\vec P_2)^2
+\frac16\vec P_1\cdot\vec P_2+
\bigg(-\frac{13}{216}-\frac{1}{9\epsilon}\bigg) q^2
-\frac{1}{24}\frac{[(\vec P_1-\vec P_2)\cdot\vec q]^2}{q^2}\bigg]\bigg\}
\bigg\rangle\nonumber \\ && + \delta\,Z^3\,\Big<\delta^d(r_1) + \delta^d(r_2)\Big>\,.
\end{eqnarray}
Here, $\delta$ is as yet undetermined state-independent coefficient, which will be obtained in the
next section by matching the hydrogenic result.

\section{Total result for $\bm{\alpha^7\,m}$ one-loop self-energy}
\label{sec:SE}

In this section we will obtain the total result for the ${\alpha^7\,m}$ one-loop self-energy
correction that is beyond the relativistic correction to the Bethe logarithm already calculated
in Ref.~\cite{patkos:20}. We add together the previously calculated parts, namely, the
second-order contribution given by Eq.~(\ref{Eseco:tot}), the low-energy contribution given by
Eq.~(\ref{LEtot2}), the middle-energy contribution given by Eq.~(\ref{eq:EM}), and the
high-energy contribution given by Eq.~(\ref{eq:EH}),
\begin{eqnarray}
E^{(7)}_{\rm SE} = E^{(7)}_{\rm sec, SE} + E_L^{\Lambda} + E_M + E_H
 = E_{\rm SE}^A + E_{\rm SE}^B\,.
\end{eqnarray}
We have split the total result into two parts, $E_{\rm SE}^A$ and $E_{\rm SE}^B$. $E_{\rm
SE}^A$ contains those two-body and three-body terms that are already in the coordinate
representation, whereas $E_{\rm SE}^B$ consists of the remaining electron-electron two-body terms
that are presently written in the momentum representation. We find that  all terms $\propto
1/\epsilon$ cancel each other in the sum, so we can make the transition $d\to 3$.
The result is still dependent on the intermediate momentum cutoff parameter $\lambda$. The
examination presented in Appendix~\ref{app:lambda} demonstrates that all $\lambda$-dependent
terms cancel when we add together $E^{(7)}_{\rm SE}$, the photon-exchange contribution derived in
Ref.~\cite{patkos:20}, and the Bethe-logarithm corrections calculated in
Ref.~\cite{yerokhin:18:betherel}. Therefore, we can just set $\lambda \to 1$ everywhere. The
resulting expression, in atomic units and with the factor $\alpha^7$ pulled out, is
\begin{eqnarray}\label{eq:setot:1}
&&E^{A}_{\rm SE}
=2\bigg\langle H''^{(5)}_{\rm SE}\frac{1}{(E_0-H_0)'}H''^{(4)}\bigg\rangle
+\frac1\pi\bigg(\frac59+\frac{1}{3}\mathcal{L}\bigg)
\bigg\langle H_R'\frac{1}{(E_0-H_0)'}H_R\bigg\rangle
+\bigg\langle\frac{1}{\pi}\bigg\{\frac{163}{240}\frac{Z^2}{r_1^4}
- \bigg(\frac{589}{360}+\frac{7\sigma_1\cdot\sigma_2}{96}\bigg)\frac{Z\vec r_1}{r_1^3}\cdot\frac{\vec r}{r^3}\nonumber\\
&&
+ \frac59 \bigg[
\bigg(E_0+\frac{Z}{r_2}-\frac{1}{r}\bigg)^2\frac{Z}{r_1}
+2\bigg(E_0+\frac{Z}{r_2}-\frac{1}{r}\bigg)\frac{Z^2}{r^2_1}
-\frac12\,p_1^2\frac{Z}{r_1}\,p_2^2 - \bigg(E_0+\frac{Z}{r_2}-\frac{1}{r}-\frac{p_2^2}{2}\bigg)\frac{Z^2}{r_1^2}
+ \frac12\,\vec p_1\,\frac{Z^2}{r_1^2}\,\vec p_1
\nonumber\\
&&+p_1^i\,\frac{Z}{r_1\,r}\bigg(\delta^{ij}+\frac{r^i r^j}{r^2}\bigg)\,p_2^j
\bigg]
+ \frac{779}{1800}\,\vec p_1\,\pi Z\delta^3(r_1)\,\vec p_1
+\mathcal{L}\bigg(\frac14\frac{Z^2}{r_1^4} + \frac13 \bigg[
\bigg(E_0+\frac{Z}{r_2}-\frac{1}{r}\bigg)^2\frac{Z}{r_1}
+2\bigg(E_0+\frac{Z}{r_2}-\frac{1}{r}\bigg)\frac{Z^2}{r^2_1}
\nonumber\\
&&
-\frac12\,p_1^2\frac{Z}{r_1}\,p_2^2 - \bigg(E_0+\frac{Z}{r_2}-\frac{1}{r}-\frac{p_2^2}{2}\bigg)\frac{Z^2}{r_1^2}
+ \frac12\,\vec p_1\,\frac{Z^2}{r_1^2}\,\vec p_1
+p_1^i\,\frac{Z}{r_1\,r}\bigg(\delta^{ij}+\frac{r^i r^j}{r^2}\bigg)\,p_2^j\bigg]
- \frac23\frac{Z\vec r_1}{r_1^3}\cdot\frac{\vec r}{r^3}
+ \frac{11}{30}\,\vec p_1\,\pi Z\delta^3(r_1)\,\vec p_1\bigg)
\nonumber\\
&&+\pi Z\delta^3(r_1)\bigg[\frac{491}{900} E_0-\frac{491}{900\,r_2} -\frac{509}{900} \frac{Z}{r_2}
+ \frac{509}{1800}p_2^2
+ \frac59\bigg\langle\frac{1}{r}\bigg\rangle
+\mathcal{L}\bigg(-\frac{E_0}{15}+\frac{1}{15\,r_2}-\frac{11}{15}\frac{Z}{r_2}+\frac{11}{30}p_2^2
+ \frac13\bigg\langle\frac{1}{r}\bigg\rangle\bigg) \bigg]
+(1\leftrightarrow2)\bigg\}\nonumber\\
&&+\frac{2E^{(4)}}{\pi}\bigg(\frac59+\frac13\mathcal{L}\bigg)\bigg(\bigg\langle\frac{1}{r}\bigg\rangle-2 E_0
\bigg)\bigg\rangle+ \delta\,Z^3\,\Big<\delta^3(r_1) + \delta^3(r_2)\Big>
 \,,\\
&&E_{\rm SE}^B =\bigg\langle
-\frac{2}{5}(\vec P_1-\vec P_2)^2 + \frac{919}{1800}q^2 + \frac{83}{90}\vec P_1\cdot\vec P_2
-\frac{4}{15}\frac{(\vec P_1\cdot\vec q)(\vec P_2\cdot\vec q)}{q^2}
+ \mathcal{L}\bigg(\frac{8}{15}q^2
+\frac43\vec P_1\cdot\vec P_2\bigg)
+\sigma_1\cdot\sigma_2\bigg(\frac{1}{24}(\vec P_1-\vec P_2)^2\nonumber\\
&&+\frac16\vec P_1\cdot\vec P_2+\bigg(\frac{43}{216}+\frac19\mathcal{L}\bigg)q^2
\bigg)
+\pi\bigg(-\frac{799}{360}-\frac{7\sigma_1\cdot\sigma_2}{48}-\frac43\mathcal{L}-\frac{4}{3}\ln q-\frac{4}{3}\ln\alpha\bigg)q\bigg\rangle
\,.\label{eq:setot:2}
\end{eqnarray}
Here, we transformed term $[(\vec P_1-\vec P_2)\cdot\vec q]^2/q^2$ into a three-photon form using
Eq.~(\ref{2ph:to:3ph}). The operators $1/r_1^4$ and $1/r_1^3$ are understood as distributions
examined in Appendix~\ref{app:singular}, so that their matrix elements are well defined.
$\mathcal{L}$ is obtained from $\ln\Lambda_\epsilon$ by dropping $1/\epsilon$ and $\ln\lambda$,
$\mathcal{L} = 2\ln(\alpha^{-2})-2\ln2\,$.

We now recall that the result (\ref{eq:setot:1}) is not complete because it contains as yet
undefined coefficient $\delta$ originating from the electron-nucleus Dirac $\delta$-function
terms omitted in our derivation. There are several sources of such terms. One of them is the
forward scattering amplitude of the three-photon exchange perturbed by the Breit Hamiltonian, the
correction to the current, and the retardation. Furthermore, such terms originate from the
singular operator $[Z^2/r_1^4]_\epsilon$.

We now proceed to obtaining the coefficient $\delta$. To this end, we first evaluate the
hydrogenic limit of Eq.~(\ref{eq:setot:1}) and compare it with the literature hydrogenic result
for the normalized difference $n^3E(nS)-E(1S)$. We should get an agreement because all terms
proportional to the electron-nucleus Dirac $\delta$-function vanish in the normalized difference.
Second, we match the hydrogenic limit of Eq.~(\ref{eq:setot:1})  with the known $1S$ hydrogenic
result and thus obtain the coefficient $\delta$ in Eq.~(\ref{eq:setot:1}).

\subsection{Restoration of the electron-nucleus Dirac $\bm{\delta}$ term}
\label{sec:hydr}

Dropping the electron-electron terms, writing $r_1\equiv r$, and omitting terms that do not
contribute to the $S$ states, we obtain the hydrogenic limit of Eq.~(\ref{eq:setot:1}) as
\begin{eqnarray}\label{hydrogen}
E^{(7)}_{\rm SE}({\rm hydr},nS) &=&
\frac1\pi\bigg(\frac59+\frac{1}{3}\mathcal{L}\bigg)
\bigg\langle H_R'\frac{1}{(E_0-H_0)'}H_R\bigg\rangle
+\bigg\langle\frac{1}{\pi}\bigg\{\bigg(\frac{163}{240}+\frac14\mathcal{L}\bigg)\frac{Z^2}{r^4}
+ \bigg(\frac59+\frac13\mathcal{L}\bigg) \bigg(-2 E_0^3+E_0\frac{Z^2}{r^2}
\\ &&
+\frac12\vec p\,\frac{Z^2}{r^2}\,\vec p
- 4\,E_0 E^{(4)}\bigg)+ \bigg(\frac{779}{1800}+\frac{11}{30}\mathcal{L}\bigg)\,\vec p\,\pi Z\delta^3(r)\,\vec p
+E_0\,\pi Z\delta^3(r)\bigg(\frac{491}{900} -\frac{1}{15}\mathcal{L}\bigg)\bigg\}
+\delta\,Z^3\,\delta^3(r)
\bigg\rangle
\,. \nonumber
\end{eqnarray}
We note that the hydrogenic limit of Eq.~(\ref{eq:setot:2}) vanishes because this expression
contains only the electron-electron terms. With help of the formulas from Appendix~\ref{app:hydr}, we
obtain  for the normalized difference
\begin{eqnarray}\label{hydrogen:diff}
\frac{\pi}{Z^6}\Big[n^3 E^{(7)}_{\rm SE}({\rm hydr},nS) - E^{(7)}_{\rm SE}({\rm hydr},1S)\Big]
&=& -\frac{16087}{5400} + \frac{263}{60n} - \frac{7583}{5400n^2}
+ \frac{163}{30}\big[\gamma+\Psi(n)-\ln n\big]
\nonumber\\
&&
+\ln\frac{\alpha^{-2}}{2} \bigg\{-\frac{103}{45}+\frac{4}{n}
-\frac{77}{45n^2}+4\big[\gamma+\Psi(n)-\ln n\big]\bigg\}\,.
\end{eqnarray}
This result agrees with Eq.~(3.43) of Ref.~\cite{jentschura:05:sese} (with the Bethe-logarithm
part omitted and with $Z$ set to 1), which indicates consistency of the derived formulas with the
hydrogen theory. The reason for setting $Z=1$ in the result of Ref.~\cite{jentschura:05:sese} is
that we are now using a different scaling in the low-energy part. In particular, in
Eq.~(\ref{21}) we define the Bethe logarithm to be rescaled by a factor of $\alpha^2$, whereas in
Ref.~\cite{jentschura:05:sese} it was rescaled by a factor of $(Z\alpha)^2$.

We will now take into account that for the $1S$ hydrogenic state, our result (\ref{eq:setot:1})
should match the one-loop self-energy part of the function $F_H$ given by Eq.~(5.116) from
Ref.~\cite{pachucki:93}, which is
\begin{equation}\label{annals}
F_H  =-\frac{121}{60} + \frac52\zeta(3) - \frac{5}{18}\pi^2 - \frac{61}{90}\ln2 - 3\ln^22
+\ln(Z\alpha)\bigg(\frac{163}{30} - 4 \ln2 - 4\ln\Lambda\bigg)
- \frac{5}{3}\ln\Lambda-\frac{22}{3}\ln2\ln\Lambda + \ln^2\Lambda\,.
\end{equation}
Here, $\Lambda$ is the intermediate momentum cutoff used in Ref.~\cite{pachucki:93}, which is the
same as the cutoff $\Lambda$ in the present work, see Eq.~(\ref{eq:23}). We now restore the
cutoff dependence of our result (\ref{hydrogen}) by shifting
\begin{eqnarray}
\mathcal{L} \to \mathcal{L}' &=& 2\ln\big[\alpha^{-2}\big]-2\ln(2\lambda)
=-2\ln\Lambda-2\ln2\,.
\end{eqnarray}
Eq.~(\ref{hydrogen}) for the $1S$ state thus becomes
\begin{equation}\label{1S}
\frac{\pi}{Z^6} E^{(7)}_{\rm SE}({\rm hydr},1S) =-\frac{7271}{1800}+\frac{221}{30}\ln2-4\ln^22+\ln\Lambda
\bigg(\frac{29}{15}-4\ln2-4\ln Z\bigg)
+\ln Z\bigg(\frac{163}{30}-4\ln2\bigg) + \delta\,.
\end{equation}
The matching condition
\begin{equation}
F_H =  \frac{\pi}{Z^6} E^{(7)}_{\rm SE}({\rm hydr},1S)
\end{equation}
leads to the following result for $\delta$,
\begin{equation}
\label{eq:150}
\delta = \frac{3641}{1800}-\frac{362}{45}\ln2+\ln^22+\frac52\zeta(3)-\frac{5}{18}\pi^2
+\ln\alpha\Big(\frac{163}{30}-4\ln2\Big)+\ln\Lambda\Big(-\frac{18}{5}-4\ln\alpha-\frac{10}{3}\ln2\Big)+\ln^2\Lambda\,.
\end{equation}

Next, we check that the cutoff dependence disappears when Eq.~(\ref{eq:150}) is combined together
with the Bethe logarithm. This is done in Appendix~\ref{app:hydr2}; the conclusion is that we can
just replace $\ln\Lambda \to \ln \alpha^2$ in the above expression. In this way, we obtain the
final result for the $\delta$ coefficient as
\begin{eqnarray}\label{eq:151}
\delta
&=&
\frac{3641}{1800}-\frac{362}{45}\ln2+\ln^22+\frac52\zeta(3)-\frac{5}{18}\pi^2
+\ln\alpha^{-2}\bigg(\frac{53}{60}+\frac{16}{3}\ln2\bigg)
-\ln^2\alpha^{-2}
\,.
\end{eqnarray}

Inserting Eq. (\ref{eq:151}) into Eq.~(\ref{eq:setot:1}), employing the explicit form of
$\mathcal L$, and using the identity $\sigma_1\cdot\sigma_2=2\,\vec\sigma_1\cdot\vec\sigma_2=2$
valid for $d=3$ and triplet states, we obtain for $E_{\rm SE}^A$ the following result
\begin{eqnarray}\label{total2}
E_{\textrm{SE}}^{A}
&=&
\bigg\langle H''^{(5)}_{\rm SE}\frac{1}{(E_0-H_0)'}H''^{(4)}\bigg\rangle
+\frac1\pi\bigg(\frac59+\frac23\ln\frac{\alpha^{-2}}{2} \bigg)
\bigg\langle H_R'\frac{1}{(E_0-H_0)'}H_R\bigg\rangle
\nonumber\\ &&
+\bigg\langle\frac{1}{\pi}\bigg\{\frac{163}{240}\frac{Z^2}{r_1^4} +
\frac59\bigg(E_0+\frac{Z}{r_2}-\frac{1}{r}\bigg)^2\frac{Z}{r_1}+\frac{10}{9}\bigg(E_0+\frac{Z}{r_2}-\frac{1}{r}\bigg)\frac{Z^2}{r^2_1}
-\frac{5}{18}\,p_1^2\frac{Z}{r_1}\,p_2^2 - \frac59\bigg(E_0+\frac{Z}{r_2}-\frac{1}{r}-\frac{p_2^2}{2}\bigg)\frac{Z^2}{r_1^2}\nonumber\\
&&+\frac{5}{18}\,\vec p_1\,\frac{Z^2}{r_1^2}\,\vec p_1
+\frac{5}{9}\,p_1^i\,\frac{Z}{r_1\,r}\bigg(\delta^{ij}+\frac{r^i r^j}{r^2}\bigg)\,p_2^j- \frac{1283}{720}\frac{Z\vec r_1\cdot\vec r}{r_1^3\,r^3}
+ \frac{779}{1800}\,\vec p_1\,\pi Z\delta^3(r_1)\,\vec p_1
+\ln\frac{\alpha^{-2}}{2} \bigg(\frac12\frac{Z^2}{r_1^4}
- \frac43\frac{Z\vec r_1\cdot\vec r}{r_1^3\,r^3}
\nonumber\\&&
+ \frac{11}{15}\,\vec p_1\,\pi Z\delta^3(r_1)\,\vec p_1
+ \frac23\bigg(E_0+\frac{Z}{r_2}-\frac{1}{r}\bigg)^2\frac{Z}{r_1}+\frac{4}{3}\bigg(E_0+\frac{Z}{r_2}-\frac{1}{r}\bigg)\frac{Z^2}{r^2_1}
-\frac13\,p_1^2\frac{Z}{r_1}\,p_2^2 - \frac23\bigg(E_0+\frac{Z}{r_2}-\frac{1}{r}-\frac{p_2^2}{2}\bigg)\frac{Z^2}{r_1^2}\nonumber\\
&&+\frac13\,\vec p_1\,\frac{Z^2}{r_1^2}\,\vec p_1
+\frac23\,p_1^i\,\frac{Z}{r_1\,r}\bigg(\delta^{ij}+\frac{r^i r^j}{r^2}\bigg)\,p_2^j\bigg)
+\pi Z\delta^3(r_1)\bigg[\frac{491}{900} E_0-\frac{491}{900\,r_2}
 -\frac{509}{900} \frac{Z}{r_2}
+ \frac{509}{1800}p_2^2
+ \frac59\bigg\langle\frac{1}{r}\bigg\rangle
+\ln\frac{\alpha^{-2}}{2} \nonumber\\
&&\times\bigg(-\frac{2E_0}{15}+\frac{2}{15\,r_2}-\frac{22}{15}\frac{Z}{r_2}+\frac{11}{15}p_2^2
+ \frac23\bigg\langle\frac{1}{r}\bigg\rangle\bigg)
 +Z^2
\bigg\{\frac{3641}{1800}-\frac{1289}{180}\ln2+\frac{16}{3}\ln^22+\frac52\zeta(3)-\frac{5}{18}\pi^2-\ln^2\frac{\alpha^{-2}}{2}
\nonumber\\
&&+\ln\frac{\alpha^{-2}}{2}\bigg(\frac{53}{60}+\frac{10}{3}\ln2\bigg)\bigg\}
\bigg]+(1\leftrightarrow2)\bigg\}
+\frac{2E^{(4)}}{\pi}\bigg(\frac59+\frac23\ln\frac{\alpha^{-2}}{2} \bigg)
\bigg(\bigg\langle\frac{1}{r}\bigg\rangle-2 E_0
\bigg)\bigg\rangle\,.
\end{eqnarray}

\subsection{Transformation of $\bm{E_{\rm SE}^B}$ into coordinate space}

The expression for $E_{\rm SE}^B$, given by Eq.~(\ref{eq:setot:2}), is written in momentum space
and needs to be transformed into the coordinate representation, to make a numerical
evaluation tractable. We first express the momenta $\vec P_1$ and $\vec P_2$ in terms of new
variables $\vec P$, $\vec p$, and $\vec q$, defined as
\begin{align}
\vec P =&\ \vec p_1+\vec p_2\,, \ \
\vec p = \frac12\big(\vec p_1-\vec p_2\big)\,, \ \ \mbox{\rm and} \ \ \vec{q} = \vec p_1{}' -\vec p_1\,.
\end{align}
We thus have
\begin{eqnarray}
\big(\vec P_1-\vec P_2\big)^2
&=&q^2+4\, \vec p\cdot\vec p \,{}'\,,
 \ \ \ \ \
\vec P_1\cdot\vec P_2
= \frac14\Big(P^2-q^2-4\, \vec p\cdot\vec p \,{}'\Big)\,,
\end{eqnarray}
\begin{eqnarray}
\frac{(\vec P_1\cdot\vec q)(\vec P_2\cdot\vec q)}{q^2}
&=&\Big(\frac14P^iP^j-p^i p'^j\Big)\frac{q^i q^j-\frac{\delta^{ij}}{3}q^2}{q^2}
-\frac14q^2+\frac{1}{12}P^2-\frac13\,\vec p\cdot\vec p \,{}'\,.
\end{eqnarray}
Furthermore, we employ the relation $\sigma_1\cdot\sigma_2 =2$ valid in $d = 3$ and take into
account that the operators $P^2\,\delta^3(r)$ and  $p^2\,\delta^3(r)$ vanish for triplet states.
Moreover, we perform the replacement $ q^2\rightarrow-2\,\vec p\cdot\vec p \,{}' $ because there
are no $q^2\,\ln q$ terms.
Performing these transformations
and using formulas for the Fourier transform from Appendix~\ref{app:Fourier}, we bring the
expression for $E_{\textrm{SE}}^{B}$ into the coordinate representation (with the overall factor
$\alpha^7$ pulled out),
\begin{eqnarray}\label{SE:coord}
E_{\textrm{SE}}^{B}&=&\bigg\langle
-\bigg(\frac{2108}{675}+\frac{196}{45}\ln\frac{\alpha^{-2}}{2} \bigg)
\vec p\,\delta^3(r)\,\vec p
-\frac{1}{60\pi}P^iP^j\frac{(\delta^{ij}r^2-3r^i r^j)}{r^5}+\frac{1}{15\pi}p^i\frac{(\delta^{ij}r^2-3r^i r^j)}{r^5}p^j\nonumber\\
&&+\frac{1}{3\pi\,r^4}\bigg(\frac{203}{15}+6\ln\frac{\alpha^{-2}}{2}-2\ln2-4\gamma-4\ln r
\bigg)\bigg\rangle\,.
\end{eqnarray}

The final result for the $\alpha^7\,m$ one-loop self-energy contribution beyond the
Bethe-logarithmic part is given by the sum of Eqs.~(\ref{total2}) and (\ref{SE:coord}).

\end{widetext}

\section{One-loop vacuum polarization}
\label{sec:VP}

We now turn to the derivation of the $\alpha^7\,m$ correction induced by the one-loop vacuum
polarization. Its calculation is much simpler than that of the self-energy. It will be convenient
to split the vacuum-polarization correction into the electron-nucleus ({\em en}) and the
electron-electron ({\em ee}) parts,
\begin{eqnarray}
E_{\rm VP}^{(7)} = E_{\rm VP}^{en} + E_{\rm VP}^{ee}\,.
\end{eqnarray}
The electron vacuum polarization modifies the photon propagator as
\begin{equation}
\frac{g_{\mu\nu}}{q_0^2-q^2}\to\frac{g_{\mu\nu}}{q_0^2-q^2}\,\big[1-\omega\big(q_0^2-q^2\big)\big]\,.
\end{equation}
Here, the function $\omega\big(q^2\big)$ is the Uehling correction defined as
\begin{eqnarray}\label{eq:159}
\omega\big(q^2\big) &=& \frac{\alpha}{\pi}\, q^2\int_4^\infty d(k^2)\frac{1}{k^2(k^2-q^2)}u(k^2)\,,
\end{eqnarray}
where
\begin{equation}
u(k^2)=\frac13\sqrt{1-\frac{4}{k^2}}\bigg(1+\frac{2}{k^2}\bigg)\,.
\end{equation}
The low-momentum expansion of Eq.~(\ref{eq:159}) is
\begin{equation}\label{Vq}
\omega\big(q^2\big)=
\frac{\alpha}{15\pi}\,q^2+\frac{\alpha}{140\pi}\,q^4+\cdots\,.
\end{equation}
In the coordinate representation and for the electron-nucleus interaction, these expansion terms
give rise to the following corrections to the Coulomb potential,
\begin{eqnarray}
\delta V^{(1)} &=& -\frac{4\,\alpha^2}{15}Z\,\delta^d(r_1) + (1\leftrightarrow2)\,,\\
\delta V^{(2)} &=& -\frac{\alpha^2}{35}\vec\nabla^2 Z\,\delta^d(r_1) + (1\leftrightarrow2)\,.
\end{eqnarray}

\subsection{Electron-nucleus vacuum polarization}

We start with the electron-nucleus part of the vacuum polarization. The corresponding correction
is represented as a sum of four parts,
\begin{equation}
E_{\rm VP}^{en} = E_{\rm sec}^{en} + E_L^{en} + E_H^{en} +
E_{\textrm{WK}}^{en}\,,
\end{equation}
where $E_{\rm sec}^{en}$ is the second-order Uehling correction, $E_L^{en}$ and $E_H^{en}$ are
the low-energy and the high-energy Uehling contributions, respectively; and
$E_{\textrm{VP}}^{\textrm{WK}}$ is the Wichman-Kroll part. The low- and the high-energy
contributions are induced by the exchanged momentum of the order $\alpha\,m$ and $m$,
respectively.

The low-energy part $E_L^{en}$ is induced by an effective operator $H_L^{en}$, $E_L^{en} = \lbr
H_L^{en} \rbr $, which is evaluated as
\begin{eqnarray}
H_L^{en} &=& \delta V^{(2)} + \frac{1}{8}\vec\nabla^2\,\delta^{(1)}V
\nonumber \\
&=&
-\frac{13\,\alpha^2}{210 }\vec\nabla^2 Z\,\delta^d(r_1) + (1\leftrightarrow2)\nonumber\\
&=&\frac{13\,\alpha^2}{105 }\bigg[2\bigg(E_0-V-\frac{p_2^2}{2}\bigg)Z\,\delta^d(r_1)
-\vec p_1\,Z\,\delta^d(r_1)\,\vec p_1\bigg]\,.
\nonumber\\
\end{eqnarray}

The second-order contribution is
\begin{eqnarray}
E_{\rm sec}^{en} &=& 2\,\Big\langle \delta V^{(1)}\,\frac{1}{(E_0-H_0)'}\,H'^{(4)}\Big\rangle\,.
\end{eqnarray}
Rewriting it as
\begin{eqnarray}
E_{\rm sec}^{en}
&=&-\frac{2\alpha}{15\pi }\,\Big\langle [\vec P,[V,\vec P]]\,\frac{1}{(E_0-H_0)'}\,H'^{(4)}\Big\rangle\,,
\end{eqnarray}
we obtain the second-order correction we encountered earlier. The result thus is
\begin{widetext}
\begin{eqnarray}
E_{\rm sec}^{en}
&=&-\frac{2\alpha}{15\pi}\,\bigg(
\Big\langle H_R'\frac{1}{(E_0-H_0)'}H_R\Big\rangle
+\Big\langle E^{(4)}\Big(\Big\langle\frac{2\alpha}{r}\Big\rangle-4 E_0\Big)
+\Big\{
\Big(2E_0-\frac{2\alpha}{r_2}+\Big\langle\frac{\alpha}{r}\Big\rangle
+\frac{(Z\alpha)^2}{2\epsilon}-2(Z\alpha)^2\Big)\,\pi Z\alpha\, \delta^3(r_1)\nonumber\\
&&
-\frac{1}{4}\frac{(Z\alpha)^2}{r_1^4}
+\bigg(E_0+\frac{Z\alpha}{r_2}-\frac{\alpha}{r}\bigg)^2\,\frac{Z\alpha}{r_1}
+2\bigg(E_0+\frac{Z\alpha}{r_2}-\frac{\alpha}{r}\bigg)\,\frac{(Z\alpha)^2}{r^2_1}
-\bigg(E_0+\frac{Z\alpha}{r_2}-\frac{\alpha}{r}-\frac{p_2^2}{2}\bigg)\frac{(Z\alpha)^2}{r_1^2}\nonumber\\
&&
+\frac12\vec p_1\frac{(Z\alpha)^2}{r_1^2}\,\vec p_1-\frac12p_1^2\,\frac{Z\alpha}{r_1}\,p_2^2
+p_1^i\,\frac{Z\alpha^2}{r_1\,r}\Big(\delta^{ij}+\frac{r^i r^j}{r^2}\Big)\,p_2^j
+(1\leftrightarrow2)\Big\}\Big\rangle
\bigg)\,.
\end{eqnarray}

The high-energy part is expressed as
\begin{eqnarray}
E_H^{en} &=& (4\pi Z\,\alpha)^3\,\phi^2(0)\int\frac{d^d q_1}{(2\pi)^d}\int\frac{d^d q_2}{(2\pi)^d}
\frac{1}{(\vec{q}_1)^4\, (\vec{q}_2)^4 \,(\vec{q}_{12})^2}\textrm{Tr}\Big[(\slashed{p}_1+1)\gamma_0(\slashed{p}_2+1)\frac{(\gamma_0+I)}{4}\Big]
\big[\omega(-\vec q_1{}^2)+\omega(-\vec q_2{}^2)+\omega(-\vec q{\,}^2_{12})\big]\,,
\nonumber \\
\end{eqnarray}
where $\vec q_{12} = \vec q_1-\vec q_2$ and $p_i = (1,\vec q_i)$. We evaluate the trace as
\begin{equation}\label{eq:169}
\textrm{Tr}\Big[(\slashed{p}_1+1)\gamma_0(\slashed{p}_2+1)\frac{(\gamma_0+I)}{4}\Big] = 4+\vec q_1\cdot\vec q_2\,.
\end{equation}
Denoting the part of $E_H$ induced by the first term in the right-hand side of Eq.~(\ref{eq:169})
as $E_{H1}$, we obtain
\begin{eqnarray}
E_{H1} &=& -\frac{\alpha(4\pi Z\,\alpha)^3}{\pi}\,4\phi^2(0)\int_4^\infty d(k^2)\int\frac{d^d q_1}{(2\pi)^d}\int\frac{d^d q_2}{(2\pi)^d}
\frac{u(k^2)}{k^2}
\frac{1}{q_1^4\, q_2^4 \,\vec q{\,}^2_{12}}\bigg(\frac{q_1^2}{k^2+q_1^2}+\frac{q_2^2}{k^2+q_2^2}
+\frac{\vec q{\,}^2_{12}}{k^2+\vec q{\,}^2_{12}}\bigg)\,.
\end{eqnarray}
We evaluate this integral with help of the following formula,
\begin{align}
  \int \frac{d^d k}{(2\,\pi)^d}\, \int \frac{d^d q}{(2\,\pi)^d}\, &\
  \frac{1}{[k^2]^{n_1}}\,\frac{1}{[(k-q)^2]^{n_2}}\,\frac{1}{[q^2+m^2]^{n_3}}
  =
  \frac{m^{2\,(d-n_1-n_2-n_3)}}{(4\,\pi)^d}\,
  \nonumber \\ & \times
  \frac{\Gamma(d/2-n_1)\,\Gamma(d/2-n_2)\, \Gamma(n_1+n_2-d/2)\, \Gamma(n_1+n_2+n_3-d)}
       {\Gamma(n_1)\, \Gamma(n_2)\, \Gamma(n_3)\, \Gamma(d/2)} \,.\label{a8}
\end{align}
In the special case of $m=0$, the integral vanishes because of the dimensional
regularization. We obtain
\begin{equation}
E_{H1} = -\frac27\alpha^4\Big< Z^3\delta^3(r_1)+Z^3\delta^3(r_2)\Big>\,.
\end{equation}
Similarly we evaluate the contribution due to the second term in the right-hand side of
Eq.~(\ref{eq:169}),
\begin{equation}
E_{H2} = \alpha^4\bigg(-\frac{32}{225}+\frac{1}{15\epsilon}\bigg)\Big< Z^3\delta^3(r_1)+Z^3\delta^3(r_2)\Big>\,.
\end{equation}

The final result for the one-loop Uehling electron-nucleus vacuum polarization (in atomic units)
is
\begin{eqnarray}
E^{en}_{\textrm{Ue}} &=& E_{\rm sec}^{en} + \langle H_L^{en}\rangle + E_H^{en}
=-\frac{2}{15\pi}\,\bigg\{
\bigg\langle H_R'\frac{1}{(E_0-H_0)'}H_R\bigg\rangle
+\bigg\langle E^{(4)}\bigg(\bigg\langle\frac{2}{r}\bigg\rangle-4 E_0\bigg)
\nonumber\\
&&+\bigg\{
\bigg(\frac{1}{7}E_0-\frac{1}{7\,r_2}-\frac{13Z}{7\,r_2}+\frac{13p_2^2}{14}+\bigg\langle\frac{1}{r}\bigg\rangle
+\frac{127}{105}Z^2-2Z^2\ln\alpha\bigg)\,\pi Z \delta^3(r_1)
-\frac14\frac{Z^2}{r_1^4}
+\bigg(E_0+\frac{Z}{r_2}-\frac{1}{r}\bigg)^2\,\frac{Z}{r_1}\nonumber\\
&&+2\bigg(E_0+\frac{Z}{r_2}-\frac{1}{r}\bigg)\,\frac{Z^2}{r^2_1}
+\frac{13}{14}\vec p_1\,\pi Z\delta^3(r_1)\,\vec p_1
-\bigg(E_0+\frac{Z}{r_2}-\frac{1}{r}-\frac{p_2^2}{2}\bigg)\frac{Z^2}{r_1^2}+\frac12\vec p_1\,\frac{Z^2}{r_1^2}\,\vec p_1\nonumber\\
&&-\frac12p_1^2\,\frac{Z}{r_1}\,p_2^2
+p_1^i\,\frac{Z}{r_1\,r}\bigg(\delta^{ij}+\frac{r^i r^j}{r^2}\bigg)\,p_2^j
+(1\leftrightarrow2)\bigg\}\bigg\rangle
\bigg\}\,.
\end{eqnarray}
The term with $\ln\alpha$ comes from rescaling of the operator $(Z\alpha)^2/r_1^4$ into atomic units.
The hydrogenic limit of this expression for $nS$-states is (with $r_1\equiv r$)
\begin{eqnarray}
E^{en}_{\textrm{Ue}}({\rm hydr},nS)
&=&-\frac{2}{15\pi}\,\bigg\{
\bigg\langle H_R'\frac{1}{(E_0-H_0)'}H_R\bigg\rangle
+\bigg\langle -4E_0E^{(4)} -2 E_0^3 +
\bigg(\frac{1}{7}E_0+\frac{127}{105}Z^2+Z^2\ln\alpha^{-2}\bigg)\,\pi Z \delta^3(r)
\nonumber\\
&&
-\frac14\frac{Z^2}{r^4}
+E_0 \frac{Z^2}{r^2}+\frac12\vec p\,\frac{Z^2}{r^2}\,\vec p\bigg\rangle
\bigg\}.
\end{eqnarray}
Using the expectation values of operators from Appendix~\ref{app:hydr}, we get
\begin{equation}
E^{en}_{\textrm{Ue}}({\rm hydr},nS) =Z^6\bigg[\frac{19}{35\,n^5} - \frac{4}{15\,n^4} - \frac{1724}{1575\,n^3} + \frac{4}{15\,n^3}\bigg(\gamma + \Psi(n) - \ln\frac{n}{2}\bigg)-\frac{2}{15n^3}\ln\big[(Z\alpha)^{-2}\big]\bigg]\,,
\end{equation}
in agreement with the known result \cite{eides:01}.
\end{widetext}

To complete our treatment of the electron-nucleus vacuum polarization, we have to include the
Wichman-Kroll correction, which is purely a Dirac-$\delta$-like type of contribution. It is given
by
\begin{equation}
E_{\textrm{WK}}^{en} = \bigg(\frac{19}{45}-\frac{\pi^2}{27}\bigg)\Big< Z^3\,\delta^3(r_1)+Z^3\,\delta^3(r_2)\Big>\,.
\end{equation}

\subsection{Electron-electron vacuum polarization}

The electron-electron part of the vacuum polarization is simpler to evaluate because the
corresponding high-energy and second-order contributions vanish for triplet states. The
low-energy part consists of two parts. The first one is due to the $q^4$ term of the expansion
(\ref{Vq}). Denoting the corresponding operator as $H_{L1}^{ee}$, we obtain
\begin{eqnarray}
H_{L1}^{ee} &=& \delta V^{(2)}
=\frac{\alpha^2}{35 }\vec\nabla_1^2 \delta^d(r) + (1\leftrightarrow2)\nonumber\\
&=&\frac{4\alpha^2}{35 }\vec p\,\delta^d(r)\,\vec p\,.
\end{eqnarray}

The remaining part of the electron-electron contribution is induced by the $q_0^2-q^2$ term of
the expansion (\ref{Vq}). For its calculation we again use the scattering amplitude approach in
the Feynman gauge, like in the derivation of $E_9$ in the one-loop self-energy calculation. In
this case the perturbed vertex is
\begin{equation}
\Gamma^\mu = -\frac{\alpha}{15\,\pi}\big(q_0^2-q^2\big)\,\gamma^\mu\equiv\alpha_{\textrm{VP}}\big(q_0^2-q^2\big)\,\gamma^\mu.
\end{equation}
Again, we pull out the factor $q_0^2-q^2$ and cancel it with the same factor in the denominator
of the photon propagator. The derivation then proceeds in the same way as in the self-energy
calculation, leading to the result
\begin{eqnarray}
U(\vec p_1,\vec p_2,\vec q) &=&\alpha_{\textrm{VP}}e^2\bigg[-1+\frac{ q^2}{4}
+\vec P_1\cdot\vec P_2+\frac{\sigma_1\cdot\sigma_2}{4\,d}\,q^2
\bigg]
\nonumber \\ &&+(1\leftrightarrow2)\,.
\end{eqnarray}
The first term here corresponds to the leading vacuum polarization correction, whereas the
remaining terms are the $\alpha^7m$ corrections. We now transform them into the coordinate
representation with help of the following transformations,
\begin{eqnarray}
\vec P_1\cdot\vec P_2 &=& \frac14\bigg(\vec P{}^2-\vec q{\,}^2-4\,\vec p\cdot\vec p\,{}'\bigg)\,, \\
\vec q{\,}^2&\rightarrow&-2\,\vec p\cdot\vec p\,{}'\,,\\
\sigma_1\cdot\sigma_2&\rightarrow& 2\,,
\end{eqnarray}
where the second and third equations are valid for triplet states. We then get
\begin{equation}
H_{L2}^{ee} = -\alpha_{\textrm{VP}} e^2\,\frac83\vec p\,\delta^3(r)\,\vec p
=\bigg(\frac{4\alpha^2}{15}\bigg)\,\frac83\vec p\,\delta^3(r)\,\vec p\,.
\end{equation}

The total electron-electron part of the $\alpha^7\,m$ vacuum polarization is the sum of
$H_{L1}^{ee}$ and $H_{L2}^{ee}$, with the result in atomic units
\begin{equation}
H_{\textrm{VP}}^{ee} = \frac{52}{63}\,\vec p\,\delta^3(r)\,\vec p\,.
\end{equation}

\subsection{Total vacuum polarization}
\begin{widetext}
Adding together the electron-nucleus and the electron-electron parts, we get the final result for
the one-loop vacuum polarization to the order $\alpha^7\,m$ in atomic units,
\begin{eqnarray}\label{VP:total}
E^{(7)}_{\textrm{VP}}
&=&-\frac{2}{15\pi}\,\bigg\{
\bigg\langle H_R'\frac{1}{(E_0-H_0)'}H_R\bigg\rangle
+\bigg\langle E^{(4)}\bigg(\bigg\langle\frac{2}{r}\bigg\rangle-4 E_0\bigg)
\nonumber\\
&&+\bigg\{
\bigg(\frac{1}{7}E_0-\frac{1}{7\,r_2}-\frac{13Z}{7\,r_2}+\frac{13p_2^2}{14}+\Big\langle\frac{1}{r}\Big\rangle
+\frac{127}{105}Z^2+Z^2\ln\alpha^{-2}\bigg)\,\pi Z \delta^3(r_1)
-\frac14\frac{Z^2}{r_1^4}
+\bigg(E_0+\frac{Z}{r_2}-\frac{1}{r}\bigg)^2\,\frac{Z}{r_1}\nonumber\\
&&+2\bigg(E_0+\frac{Z}{r_2}-\frac{1}{r}\bigg)\,\frac{Z^2}{r^2_1}
+\frac{13}{14}\vec p_1\,\pi Z\delta^3(r_1)\,\vec p_1
-\bigg(E_0+\frac{Z}{r_2}-\frac{1}{r}-\frac{p_2^2}{2}\bigg)\frac{Z^2}{r_1^2}+\frac12\vec p_1\,\frac{Z^2}{r_1^2}\,\vec p_1-\frac12p_1^2\,\frac{Z}{r_1}\,p_2^2
\nonumber\\
&&+p_1^i\,\frac{Z}{r_1\,r}\Big(\delta^{ij}+\frac{r^i r^j}{r^2}\Big)\,p_2^j
+(1\leftrightarrow2)\bigg\}\bigg\rangle
\bigg\}
+\bigg\langle\frac{52}{63}\vec p\,\delta^3(r)\,\vec p
+\bigg(\frac{19}{45}-\frac{\pi^2}{27}\bigg) \big[Z^3\,\delta^3(r_1)+Z^3\,\delta^3(r_2)\big]\bigg\rangle\,.
\end{eqnarray}

\end{widetext}

\section{Two-loop and three-loop contributions}
\label{sec:multiloop}

The two-loop radiative contribution is proportional to the electron-nucleus Dirac $\delta$
function and is obtained immediately from the hydrogenic result,
\begin{align}\label{eq:QED2}
E^{(7)}_{\rm rad2} = \frac{Z^2}{\pi}\,\big< \delta^{3}(r_1)+\delta^{3}(r_2)\big>\, B_{50}\,,
\end{align}
where the coefficient $B_{50}$ is known only numerically
\cite{pachucki:94,eides:95:pra,dowling:10}, $B_{50} = -21.554\,47\,(13)$.

After dropping the part contributing to the fine structure, the three-loop radiative correction
becomes also proportional to the electron-nucleus Dirac $\delta$ function and is given by
\cite{laporta:96,melnikov:00}
\begin{align}\label{eq:QED3}
E^{(7)}_{\rm rad3} = &\ \frac{Z}{\pi^2}\,\big< \delta^{3}(r_1)+\delta^{3}(r_2)\big>\,
\bigg[ -{{568\,{\rm a_4}}\over{9}}+{{85\,\zeta(5)}\over{24}}
 \nonumber\\&
-{{121\,\pi^{2}\,\zeta(3)}\over{72}} -{{84\,071\,\zeta(3)}\over{2304}} -{{71\,\ln ^{4}2}\over{27}}
 \nonumber\\&
 -{{239\,\pi^{2}\,\ln^{2}2}\over{135}} +{{4787\,\pi^{2}\,\ln 2}\over{108}}
+{{1591\,\pi^{4}}\over{3240}}
 \nonumber\\&
-{{252\,251\,\pi^{2}}\over{9720}}+{679\,441\over93\,312}
\bigg]\,,
\end{align}
where $\rm a_4 = \sum_{n=1}^\infty 1/(2^n\,n^4) = 0.517\,479\,061\dots$.

\section{Summary}

In this work we derived the radiative $\alpha^7m$ QED correction for the triplet states of a
two-electron atom. This correction consists of the one-loop self-energy part
$E_{\textrm{SE}}^{(7)}$ given by the sum of Eqs.~(\ref{total2}) and (\ref{SE:coord}), the
one-loop vacuum-polarization part $E_{\textrm{VP}}^{(7)}$ given by Eq.~(\ref{VP:total}), the
two-loop part $E_{\textrm{rad2}}^{(7)}$ given by Eq.~(\ref{eq:QED2}), and by the three-loop part
$E_{\textrm{rad3}}^{(7)}$ given by Eq.~(\ref{eq:QED3}).

In order to obtain the complete $\alpha^7m$ QED correction, we have to add to the above mentioned
contributions the nonradiative photon exchange part $E_{\textrm{exch}}^{(7)}$ derived in
Ref.~\cite{patkos:20} and the relativistic correction to the Bethe logarithm $E_{L}^{(7)}$ from
Ref.~\cite{yerokhin:18:betherel}. More specifically, the photon-exchange correction is
\begin{align}
E_{\textrm{exch}}^{(7)} = \lbr H_{\rm{exch}}^{(7)}\rbr + 2
\Big< H^{(4)} \frac1{(E_0-H_0)'}\, H^{(5)}_{\rm exch}\Big>\,,
\end{align}
where $H_{\rm{exch}}^{(7)}$ and $H_{\rm{exch}}^{(5)}$ are given,  correspondingly,  by Eqs.~(156)
and (10) of Ref.~\cite{patkos:20}, and the Bethe-logarithm correction is
\begin{align}
E_{L}^{(7)} = E_{L1}^{(7)} + E_{L2}^{(7)}+E_{L3}^{(7)}\,,
\end{align}
with $E_{L1}^{(7)} $, $E_{L2}^{(7)}$, and $E_{L3}^{(7)}$ given by Eqs.~(14), (20), and (27) of
Ref.~\cite{yerokhin:18:betherel}, respectively.

The final step of our project will be the numerical evaluation of all $\alpha^7m$ QED
corrections. The most complicated part of the computation is already accomplished in
Ref.~\cite{yerokhin:18:betherel}, where we obtained numerical results for the relativistic
corrections to the Bethe logarithm for the $2\,^3\!S$ and $2\,^3\!P$ states. A computation of the
remaining photon-exchange and radiative contributions looks relatively straightforward. The only
new feature as compared to our previous calculations of higher-order QED corrections in helium
\cite{patkos:16:triplet,patkos:17:singlet} is the appearance of singular operators with $\ln r$,
such as $\ln r/r^4$ in Eq.~(\ref{SE:coord}). Matrix elements of such operators should be
understood in terms of a special limit as discussed in
Appendix~\ref{app:singular}. In particular, the regularized operator $\ln r/r^4$ is defined by
Eq.~(\ref{eq:E3}). We are currently working on developing an effective computational scheme for
such operators.

\begin{acknowledgments}
K.P. and V.P. acknowledge support from the National Science Center (Poland) Grant No.
2017/27/B/ST2/02459. V.P. was also supported by the Czech Science Foundation - GA\v{C}R (Grant
No. P209/18-00918S).
\end{acknowledgments}


\appendix

\section{Electromagnetic form factors}
\label{app:ffactors}

The electromagnetic form factors $F_1$ and $F_2$ of an electron are defined as
\begin{eqnarray}
\gamma_\mu \to \Gamma_\mu &=&
\gamma_\mu + \gamma_\mu\,F_1\big(q_0^2-q^2\big)\nonumber\\&& + \frac{{\rm i}}{2m} F_2\big(q_0^2-q^2\big)
\left( {{\rm i}\over 2} \right)[\gamma_\mu,\qsla]\,,
\label{A1}
\end{eqnarray}
where $q$ is the outgoing photon momentum. The small-$q$ expansion of the one-loop formfactors in
$D=4-2\epsilon$ is \cite{bonciani:03}
\begin{eqnarray}
F_1^{(1)}\big(q^2\big) &=& \frac{\alpha}{\pi}\,\biggl[q^2 \left( - {1 \over 8}
- {1 \over 6\epsilon} - {1 \over 2} \epsilon \right)
 \nonumber\\ &&
+ q^4 \left( - {11 \over 240} - {1 \over 40\epsilon}
- {5 \over 48} \epsilon \right)\biggr]\,,\nonumber\\ &&\\
F_2^{(1)}\big(q^2\big) &=&  \frac{\alpha}{\pi}\,\biggl[
{1 \over 2} + 2 \epsilon
+ q^2 \left( {1 \over 12} + {5 \over 12} \epsilon  \right)
 \nonumber\\ &&
+ q^4 \left( {1 \over 60} + {11 \over 120} \epsilon  \right)\biggr]\,.
\end{eqnarray}
\section{Integrations in spheroidal coordinates in $\bm{d}$ dimensions}
\label{app:spheroid}

To introduce the spheroidal coordinates in $d$ dimensions, we start with the volume element in
the spherical coordinates in $d-1$ dimensions, with variable $z$ for the last dimension,
\begin{equation}
dV = r^{d-2} dr \,dz \,\Omega_{d-1}\,.
\end{equation}
Let us  define
\begin{eqnarray}
r_1 &=& \sqrt{r^2+\left(z+\frac{a}{2}\right)^2},\\
r_2 &=& \sqrt{r^2+\left(z-\frac{a}{2}\right)^2},
\end{eqnarray}
and introduce spheroidal variables $\xi$ and $\eta$ as
\begin{eqnarray}
\xi &=&\frac{r_1+r_2}{a},\\
\eta &=&\frac{r_1-r_2}{a}.
\end{eqnarray}
The following relations hold
\begin{eqnarray}
z &=& \frac{\xi\,\eta}{2},\\
r^2 &=& \frac14 (\xi^2-1)(1-\eta^2).
\end{eqnarray}
In the new coordinates, the volume element is
\begin{eqnarray}
dV &=&  \Bigl(\frac{a}{2}\Bigr)^d \,\Omega_{d-1}\big[(\xi^2-1)(1-\eta^2)\big]^\frac{d-3}{2}\nonumber\\
&&\times\,(\xi-\eta)(\xi+\eta)\, d\xi\,d\eta\,,
\end{eqnarray}
and $\eta\in(-1,1)$, $\xi\in(1,\infty)$.

We now consider the integral of the form
\begin{align}
&\ \int d^d k \,f\biggl(\left|\vec k+\frac{\vec a}{2}\right|,  \left|\vec k-\frac{\vec a}{2}\right| \biggr)\nonumber \\
=&\ \int d^d k\,f(r_1,  r_2)\nonumber \\
=&\ \Bigl(\frac{a}{2}\Bigr)^d \,\Omega_{d-1}\int d\xi\,d\eta\,\big[(\xi^2-1)(1-\eta^2)\big]^\frac{d-3}{2}\nonumber \\
&\ \times(\xi-\eta)(\xi+\eta)\, f\biggl(a\,\frac{(\xi+\eta)}{2},  a\,\frac{(\xi-\eta)}{2}\biggr)\,.
\end{align}
In spheroidal coordinates, it is just a two-dimensional integral over $\xi$ and $\eta$.

The particular case of such an integral with integer powers $i$ and $j$
\begin{equation}
J_{ij} = \int\ d^d k\,\frac{1}{\left|\vec k+\frac{\vec q}{2}\right|^i}
\frac{1}{\left|\vec k-\frac{\vec q}{2}\right|^j}\frac{1}{\left|\vec k+\frac{\vec q}{2}\right|+\left|\vec k-\frac{\vec q}{2}\right|}
\end{equation}
can be transformed to the spheroidal coordinates as
\begin{align}
J_{ij} = &\  q^{d-i-j-1}\,2^{i+j-d}\Omega_{d-1}\int_1^{\infty} d\xi\, \int_{-1}^{1}d\eta \\
&\times \,\big[(\xi^2-1)(1-\eta^2)\big]^\frac{d-3}{2}\,
(\xi-\eta)^{1-j}(\xi+\eta)^{1-i}\,\frac{1}{\xi}\,. \nonumber
\end{align}
The integrations over $\xi$ and $\eta$ can now be performed for each particular $i$ and $j$, yielding results in agreement with
those from Appendix~C of Ref.~\cite{patkos:20}. The advantage of using the spheroidal
coordinates, however, is that this approach can be applied also to non-integer values of $i$ and
$j$, in particular, for the case when they are equal to $d$, which is needed in the evaluation of
the middle-energy contribution.

\section{Cancelation of the $\bm{\lambda}$ dependence}
\label{app:lambda}
In this section we demonstrate that the sum of all $\alpha^7\,m$ terms depending on the
intermediate momentum cutoff $\lambda$ vanishes. The dimensionless cutoff parameter $\lambda$
appears when the integral over the photon momentum $k$ is divided into the $k< \Lambda$ and $k>
\Lambda$ regions, $\Lambda = \alpha^2 \lambda$, see  Eq.~(\ref{eq:23}). In order to cancel the
$\lambda$-dependent terms, we need to add the radiative contribution calculated in this
work, the non-radiative photon-exchange contribution from Ref.~\cite{patkos:20}, and the
Bethe-logarithm correction from Ref.~\cite{yerokhin:18:betherel}.

\begin{widetext}
We first address the $\lambda$-dependent part of the radiative correction from
Eqs.~(\ref{Eseco:tot}) and (\ref{LEtot2}). Denoting it as $E_{\rm{rad}}^\lambda$, we have
\begin{eqnarray}\label{D1}
E_{\rm{rad}}^\lambda
&=&\ln\lambda\bigg\{-\frac{2}{3\pi}
\bigg\langle H_R'\frac{1}{(E_0-H_0)'}H_R\bigg\rangle+
\bigg\langle\phi\bigg|\frac{1}{\pi}
\bigg\{\pi Z\delta^{d}(r_1)\bigg[
-\frac{2}{15}\bigg(-E_0+\frac{1-11Z}{r_2}+\frac{11}{2}p_2^2\bigg)
-\frac{2}{3}\bigg\langle\frac{1}{r}\bigg\rangle
\bigg]\nonumber\\
&&-\frac{1}{2}\frac{Z^2}{r_1^4}
+\frac{5}{3}\frac{Z\vec r_1}{r_1^3}\cdot\frac{\vec r}{r^3}
-\frac{11}{15}p_1^i\,\pi Z\delta^{d}(r_1)\,p_1^i
-\frac23\bigg(E_0+\frac{Z}{r_2}-\frac{1}{r}\bigg)^2\frac{Z}{r_1}
-\frac43\bigg(E_0+\frac{Z}{r_2}-\frac{1}{r}\bigg)\frac{Z^2}{r^2_1}\nonumber\\
&&
+\frac13\,p_1^2\frac{Z}{r_1}\,p_2^2+\frac43 E_0\,E^{(4)}
-\frac23E^{(4)}\bigg\langle\frac{1}{r}\bigg\rangle +\frac23 Y_1
-\frac23\,p_1^i\,\frac{Z}{r_1\,r}\bigg(\delta^{ij}+\frac{r^i r^j}{r^2}\bigg)\,p_2^j
+(1\leftrightarrow2)\bigg\}\nonumber\\
&&-
\frac{8}{3}\vec P_1\cdot\vec P_2-\frac{16}{15} q^2
-\frac{4}{3}\frac{\big[(\vec P_1-\vec P_2)\cdot\vec q\big]^2}{q^2}
-\frac{2q^2}{9}\sigma_1\cdot\sigma_2+\frac{10\pi}{3}q
\bigg|\phi\bigg\rangle\bigg\}\,.
\end{eqnarray}
Second, we need to account for the $\lambda$-dependent part of the non-radiative photon-exchange
correction derived in our previous paper \cite{patkos:20}. It is
\begin{eqnarray}
E_{\rm{exch}}^\lambda&=&\ln\lambda\,\bigg\langle
\bigg\{
\frac{8}{15}\bigg[p_1^i,\bigg[\frac{Z}{r_1},p_1^j\bigg]\bigg]
\bigg(-4\delta^{ij}+\frac{q^i q^j}{q^2}\bigg)\frac{1}{q^2}
-\frac{7}{15\pi}
\frac{Z\vec r_1}{r_1^3}\cdot\frac{\vec r}{r^3}
+(1\leftrightarrow2)\bigg\}+
\frac45(\vec P_1-\vec P_2)^2\nonumber\\
&&+\frac{16}{15}q^2
+\frac83\vec P_1\cdot\vec P_2
+\frac43\frac{[(\vec P_1-\vec P_2)\cdot\vec q]^2}{q^2}
+\frac{2q^2}{9}\sigma_1\cdot\sigma_2 -\frac{6\pi}{5}q
\bigg\rangle\,.
\end{eqnarray}
It is advantageous to use this form of the expression in the momentum representation and not the
final formula in the coordinate representation presented in Ref.~\cite{patkos:20} because of
strong cancellation of the electron-electron terms in the sum with Eq.~(\ref{D1}). Adding
the two contributions
and then transforming this result into the coordinate representation with help of formulas from
Appendix~\ref{app:Fourier}
we obtain for $E^\lambda = E^\lambda_{\rm{exch}}+E^\lambda_{\rm{rad}}$ the result
\begin{eqnarray}
E^\lambda
&=&\ln\lambda\bigg\{-\frac{2}{3\pi}
\bigg\langle H_R'\frac{1}{(E_0-H_0)'}H_R\bigg\rangle+
\bigg\langle\phi\bigg|\frac{1}{\pi}
\bigg\{\pi Z\delta^{d}(r_1)\bigg[
-\frac{2}{15}\bigg(-E_0-\frac{41}{3r_2}-\frac{11Z}{r_2}+\frac{11}{2}p_2^2\bigg)\nonumber\\
&&
-\frac{2}{3}\bigg\langle\frac{1}{r}\bigg\rangle
\bigg]-\frac{1}{2}\frac{Z^2}{r_1^4}
+\frac{6}{5}\frac{Z\vec r_1\cdot\vec r}{r_1^3 r^3}
-\frac{11}{15}p_1^i\,\pi Z\delta^{d}(r_1)\,p_1^i
-\frac23\bigg(E_0+\frac{Z}{r_2}-\frac{1}{r}\bigg)^2\frac{Z}{r_1}
-\frac43\bigg(E_0+\frac{Z}{r_2}-\frac{1}{r}\bigg)\frac{Z^2}{r^2_1}\nonumber\\
&&
+\frac13\,p_1^2\frac{Z}{r_1}\,p_2^2+\frac43 E_0\,E^{(4)}
-\frac23E^{(4)}\bigg\langle\frac{1}{r}\bigg\rangle +\frac23 Y_1
-\frac23\,p_1^i\,\frac{Z}{r_1\,r}\bigg(\delta^{ij}+\frac{r^i r^j}{r^2}\bigg)\,p_2^j
+\frac{Z}{15}\frac{\big(\delta^{ij}r_1^2-3\,r_1^i r_1^j\big)}{r_1^3}\frac{r^i r^j}{r^3}\nonumber\\
&&
+(1\leftrightarrow2)\bigg\}
+\frac{8}{5}\vec p\,\delta^d(r)\,\vec p-\frac{32}{15\pi}\frac{1}{r^4}
\bigg|\phi\bigg\rangle\bigg\}\,.
\end{eqnarray}
$E^\lambda$ has to be combined with the $\lambda$-dependent part of the relativistic correction
to the Bethe logarithm, which will be denoted as $E_{\rm{Bethe}}^\lambda$. It is given by the sum of
the contributions induced by the asymptotic coefficients given in Eqs.~(A7), (A10), (B7), (B9), (C21)
and (C24) of Ref.~\cite{yerokhin:18:betherel}, which is
\begin{eqnarray}
E_{\rm{Bethe}}^\lambda
&=&\ln\lambda\bigg\{\frac{2}{3\pi}
\bigg\langle H_R'\frac{1}{(E_0-H_0)'}H_R\bigg\rangle+
\bigg\langle\phi\bigg|\frac{1}{\pi}
\bigg\{\pi Z\delta^{d}(r_1)\bigg[
-\frac{2}{15}\bigg(E_0+\frac{41}{3r_2}+\frac{11Z}{r_2}-\frac{11}{2}p_2^2\bigg)\nonumber\\
&&
+\frac{2}{3}\bigg\langle\frac{1}{r}\bigg\rangle
\bigg]-\frac{6}{5}\frac{Z\vec r_1\cdot\vec r}{r_1^3 r^3}
+\frac{11}{15}p_1^i\,\pi Z\delta^{d}(r_1)\,p_1^i
+\frac23\bigg(E_0+\frac{Z}{r_2}-\frac{1}{r}\bigg)^2\frac{Z}{r_1}
+\frac43\bigg(E_0+\frac{Z}{r_2}-\frac{1}{r}\bigg)\frac{Z^2}{r^2_1}
-\frac{Z^3}{r_1^3}\nonumber\\
&&
-\frac13\,p_1^2\frac{Z}{r_1}\,p_2^2-\frac43 E_0\,E^{(4)}
+\frac23E^{(4)}\bigg\langle\frac{1}{r}\bigg\rangle -\frac53 Y_1
+\frac23\,p_1^i\,\frac{Z}{r_1\,r}\bigg(\delta^{ij}+\frac{r^i r^j}{r^2}\bigg)\,p_2^j
-\frac{Z}{15}\frac{\big(\delta^{ij}r_1^2-3\,r_1^i r_1^j\big)}{r_1^3}
\frac{r^i r^j}{r^3}\nonumber\\
&&
+(1\leftrightarrow2)\bigg\}
-\frac{8}{5}\vec p\,\delta^d(r)\,\vec p+\frac{32}{15\pi}\frac{1}{r^4}
\bigg|\phi\bigg\rangle\bigg\}\,.
\end{eqnarray}
\end{widetext}
We now take into account the identity
\begin{equation}
\frac{Z^2}{r_1^4} = -2\frac{Z^3}{r_1^3} -2 Y_1 - 12 Z^3\pi\,\delta^d(r_1)\,,
\end{equation}
where we should drop the second term on the right-hand side, in accordance with the procedure of
omitting all the electron-nucleus  $\delta$-like contributions at this stage of the derivation.
Therefore, we find that $E^\lambda+ E_{\rm{Bethe}}^\lambda = 0$.

The cancellation of the $\lambda$-dependent terms  proportional to the pure electron-nucleus Dirac
$\delta$-function is demonstrated in Appendix~\ref{app:hydr2}. After we checked that all
$\lambda$-dependent terms vanish as they should, we can just set $\lambda \to 1$ in all final
formulas.

\section{Cancelation of the $\bm{\lambda}$ dependence in the hydrogenic limit} \label{app:hydr2}

In Appendix~\ref{app:lambda} we proved that all $\lambda$-dependent $\alpha^7\,m$ terms vanish,
with the exception of pure electron-nucleus Dirac-$\delta$ contributions which were omitted in
the derivation. The $\delta$-like contribution was restored in Sec.~\ref{sec:hydr} by matching
our results against the known hydrogenic limit. Here we will show that the $\lambda$-dependent
terms proportional to the electron-nucleus Dirac-$\delta$ function in the hydrogenic limit vanish
as well.

Let us return to the $\lambda$-dependent part of the relativistic correction to the Bethe logarithm
from Ref.~\cite{yerokhin:18:betherel}, now keeping the $\delta$-like terms. Performing the
hydrogenic limit and taking only terms contributing to $S$ states, we have
\begin{align}
E_{\rm{Bethe}}^\lambda &\, ({\rm hydr})
 =\ln\lambda \, \Bigg\{ \frac{2}{3\pi}
\bigg\langle H_R'\frac{1}{(E_0-H_0)'}H_R\bigg\rangle
 \nonumber \\ &
+ \bigg\langle\phi\bigg|\frac{1}{\pi} \bigg\{\pi Z\delta^{d}(r)\bigg[Z^2\bigg(
-\frac{12}{5}+\frac{10}{3}\ln2-\ln\lambda\bigg)
 \nonumber\\&
-\frac{2}{15}E_0 \bigg] +\frac{11}{15}p^i\,\pi Z\delta^{d}(r)\,p^i -\frac43 E_0^3-\frac13
E_0\frac{Z^2}{r^2}
 \nonumber\\&
-\frac{Z^3}{r^3}+\frac56\vec p\,\frac{Z^2}{r^2}\,\vec p -\frac83
E_0\,E^{(4)}\bigg\} \bigg|\phi\bigg\rangle\Bigg\}\,.
\end{align}
Using results for the matrix elements from Appendix~\ref{app:hydr}, we obtain for the $1S$ state,
\begin{equation}\label{F2}
E_{\rm{Bethe}}^\lambda({\rm hydr},1S) = -\ln^2\lambda + \ln\lambda\bigg(\frac{5}{3}+\frac{22}{3}\ln2\bigg)\,.
\end{equation}

Switching to the same cutoff $\Lambda$ as in $F_H$ from Eq.~(\ref{annals}), $\Lambda =
\alpha^2\lambda$, we see that all cutoff-dependent terms cancel in the sum
$E_{\rm{Bethe}}^\lambda({\rm hydr},1S) + F_H$.

\section{Expectation values of singular operators} \label{app:singular}

In our derivation we encounter expectation values of several singular operators, which need to be
evaluated in the coordinate representation. These expectation values should be understood in the
sense of a distribution. Specifically, the expectation values of the operators $1/r^3$, $1/r^4$,
and $\ln r/r^4$ are defined by the following limits
\begin{align} \label{1/r^3}
\bigg\langle\phi\bigg| \frac{1}{r^3}\bigg|\psi\bigg\rangle
\equiv&\ \lim_{a\rightarrow0}\int d^3r\,\phi^*(\vec r)\psi(\vec r)\nonumber\\
&\ \times\bigg[\frac{1}{r^3}\Theta(r-a)+
4\,\pi\,\delta^3(r)(\gamma+\ln a)\bigg]\nonumber\\
=& \lim_{a\rightarrow0}\int_a^\infty d r \, \frac{f(r)}{r} + f(0)\,(\gamma+\ln a)\,,
\end{align}

\begin{align}
\bigg\langle\phi\bigg| \frac{1}{r^4}\bigg|\psi\bigg\rangle
\equiv& \lim_{a\rightarrow0}\int_a^\infty d r \, \frac{f(r)}{r^2} -\frac{f(0)}{a}+ f'(0)\,(\gamma+\ln a)\,,
\end{align}

\begin{align}
\bigg\langle\phi\bigg| \frac{\ln r}{r^4}\bigg|\psi\bigg\rangle
\equiv& \lim_{a\rightarrow0}\int_a^\infty d r \, \frac{f(r)\,\ln r}{r^2} - f(0)\,\frac{(1+\ln a)}{a}
\nonumber \\ &\ + f'(0)\,\frac{\ln^2 a}{2}\,,\label{eq:E3}
\end{align}
where
\begin{align}
f(r) = \int d\Omega\, \phi^*(\vec r)\,\psi(\vec r)\,.
\end{align}

\section{Hydrogenic expectation values} \label{app:hydr}

Here we list the expectation values of various $\alpha^7\,m$ operators for the hydrogenic $S$
states. They are:
\begin{align}
&E_0 =-\frac{Z^2}{2n^2}\,,\\
&E^{(4)} = Z^4\bigg(\frac{3}{8n^4} - \frac{1}{2n^3}\bigg)\,,\\
&\frac{Z^2}{r^4} =\frac{8Z^6}{n^3} \bigg(-\frac53+ \frac{1}{2n}+\frac{1}{6n^2}+\gamma+\Psi(n)-\ln \frac{n}{2Z}\bigg)\,,
\nonumber \\ \\
&\frac{Z^3}{r^3} =\frac{4Z^6}{n^3} \bigg(\frac12- \frac{1}{2n}-\gamma-\Psi(n)+\ln \frac{n}{2Z}\bigg)\,,\\
&\frac{Z^2}{r^2} =\frac{2Z^4}{n^3}\,,\\
&\vec p\,\frac{Z^2}{r^2}\,\vec p =Z^6\bigg(-\frac{2}{3n^5} + \frac{8}{3n^3}\bigg)\,,\\
&\vec p\,\pi Z\delta^d(r)\,\vec p =0\,,\\
&\pi Z\delta^d(r_1) =\frac{Z^4}{n^3}\,,
\end{align}
where $\Psi(n) = \Gamma'(n)/\Gamma(n)$. The expectation value of $Z^3/r^3$ and $Z^2/r^4$ were calculated
according to the definitions in Appendix \ref{app:singular}.
Note that the terms with $\ln Z$ originate from the rescaling $r\rightarrow Z^{-1}\, r$ which was
needed for the correct matching of our results with the hydrogenic limit, see discussion under
Eq.~(\ref{hydrogen:diff}).

The hydrogenic limit of the second-order correction is
\begin{eqnarray}
\bigg\langle H_R'\frac{1}{(E_0-H_0)'}H_R\bigg\rangle &=& Z^6\bigg(-\frac{1}{n^6}-\frac{4}{3n^5}+\frac{3}{n^4}+\frac{7}{3n^3}\bigg)\,,
\nonumber \\
\end{eqnarray}
where the operators $H_R$ and $H_R'$ act on ket-states as
\begin{eqnarray}
H_R|\phi\rangle &=&\bigg[-\frac12\bigg(E_0+\frac{Z}{r}\bigg)^2-\frac{Z}{4}\frac{\vec r\cdot\vec\nabla}{r^3}
\bigg]|\phi\rangle\,,\\
H_R'|\phi\rangle &=& -2Z\frac{\vec r\cdot\vec\nabla}{r^3}|\phi\rangle\,.
\end{eqnarray}

\section{Fourier transform}
\label{app:Fourier}

Here we list the formulas needed to transform our formulas from momentum space into the
coordinate representation. The results are \cite{patkos:20}
\begin{align}
\int\frac{d^3q}{(2\pi)^3} \,e^{{\rm i}\vec q\cdot\vec r}\,\frac{4\pi}{q^2}
=&\,\frac{1}{r}\,,\\
\int\frac{d^3q}{(2\pi)^3} \,e^{{\rm i}\vec q\cdot\vec r}\,4\pi\frac{q^i}{q^2}
=&\,{\rm i}\frac{r^i}{r^3}\,,\\
\int\frac{d^3q}{(2\pi)^3} \,e^{{\rm i}\vec q\cdot\vec r}\,4\pi \frac{q^i q^j-\frac{\delta^{ij}}{3}q^2}{q^2}
=&\,\frac{\delta^{ij}r^2-3r^i r^j}{r^5}\,,\\
\int\frac{d^3q}{(2\pi)^3} \,e^{{\rm i}\vec q\cdot\vec r}\,4\pi\,\frac{q^i q^j}{q^4}
=&\,\frac{1}{2r^3}\big(\delta^{ij}r^2-r^ir^j\big)\,.
\end{align}
The transformation of singular operators $1/r^4$ and $\ln r/r^4$ is more complicated. Here we
present results valid for triplet states \cite{patkos:20}
\begin{eqnarray}
\int d^3r\,e^{{\rm i}\vec q\cdot\vec r}\,\frac{1}{r^4}  &=& \lim_{\varepsilon\rightarrow0}
\int d^3r\,e^{{\rm i}\vec q\cdot\vec r}\bigg[\frac{1}{r^4}\theta(r-\varepsilon)-
4\pi \delta^3(r)\,\frac{1}{\varepsilon}\bigg]\nonumber\\
&=& \lim_{\varepsilon\rightarrow0}
2\pi\int_{-1}^1 dx\int_\varepsilon^\infty\frac{dr}{r^2}\,e^{{\rm i} q r x} - \frac{4\pi}{\varepsilon}
\nonumber\\
&=&
-\pi^2\,q\,.
\end{eqnarray}
and
\begin{eqnarray}\label{eq:G6}
&&\int d^3r\,e^{{\rm i}\vec q\cdot\vec r}\,\frac{\ln r}{r^4}
 \nonumber \\
&=& \lim_{\varepsilon\rightarrow0} \int d^3r\,e^{{\rm i}\vec q\cdot\vec r}\bigg[\frac{\ln
r}{r^4}\,\theta(r-\varepsilon)-
4\pi \delta^3(r)\,\frac{1+\ln\varepsilon}{\varepsilon}\bigg]\nonumber\\
&=& \lim_{\varepsilon\rightarrow0}
2\pi\int_{-1}^1 dx\int_\varepsilon^\infty\frac{dr}{r^2}\ln r\,e^{{\rm i} q r x} - 4\pi\,\frac{1+\ln\varepsilon}{\varepsilon}
\nonumber\\
&=&
 \pi^2\bigg(-\frac32+\gamma+\ln q\bigg)\,q\,.
\end{eqnarray}

\end{document}